\documentclass[preprint,aip,jmp,longbibliography,groupedaddress,]{revtex4-1}
\usepackage{amsmath}
\usepackage{amsfonts}
\usepackage{amssymb}
\usepackage{hyperref}
\usepackage{appendix}
\usepackage{amsmath,mathtools}
\usepackage{array}
\usepackage{mathtools}
\usepackage{makecell}
\usepackage{empheq}
\usepackage{titlesec}
\newtheorem{remark}{Remark}
\newtheorem{theorem}{Theorem}

\begin{document}
	\title{ \Large  Fifth-order superintergrable quantum systems separating in Cartesian coordinates. Doubly exotic potentials} 
	\author{Ismail Abouamal}
	\email[]{abouamali@dms.umontreal.ca}
	\author{Pavel Winternitz}
	\email[]{wintern@crm.umontreal.ca}
	\affiliation{D\'epartement de Math\'ematiques et de Statistiques and Centre de Recherche Math\'ematiques, Universit\'e de Montr\'eal, C.P.6128, Succursale Centre-Ville, Montr\'eal, Qu\'ebec, H3C 3J7, Canada}
	\date{\today}
	\begin{abstract}
		\noindent We consider a two dimensional quantum Hamiltonian separable in Cartesian coordinates and allowing a fifth-order integral of motion. We impose the superintegrablity condition and find all doubly exotic superintegrable potentials(i.e potentials $V(x,y)=V_1(x)+V_2(y)$ where neither $V_1(x)$ nor $V_2(y)$ satisfy a linear ODE) allowing the existence of such an integral. All of these potentials are found to have the Painlev\'e property. Most of them are expressed in terms of known Painlev\'e transcendents or elliptic functions but some may represent new higher order Painlev\'e transcendents.
	\end{abstract}
	\maketitle 
	\section{Introduction}
	We consider a quantum superintegrable Hamiltonian system in two-dimensional space $E_2$ with two integrals of motion (in addition to the Hamiltonian), namely
	\begin{equation}
	\mathcal{H}= p_1 ^2+ p_2^2+V(x,y),\ \ p_1=-i \hbar \frac{\partial}{\partial x}, \ \  p_2=-i \hbar \frac{\partial}{\partial y} , \label{Hamiltonian}
	\end{equation}
	\begin{equation}
	Y= p_1 ^2- p_2^2+g(x,y),\label{PW-integral5-I}
	\end{equation}
	\begin{equation}
	X=\frac{1}{2}\sum \limits_{l=0}^{[\frac{N}{2}]}\sum \limits_{j=0}^{N-2l}\{f_{j,2l}(x,y), p_1^j p_2^{N-2l-j}\}. \label{integral5-I}	
	\end{equation}
Above the notation is $\{A,B\}=AB+BA$ and in this article we take $N=5$. The functions $V(x,y),g(x,y)$ and $f_{j,2l}(x,y)$ are to be determined from the conditions 
\begin{equation}
[\mathcal{H},X]=0, \ \ \ [\mathcal{H},Y]=0.
\end{equation}
The existence of the second order integral $Y$ implies that the system is integrable and separable in Cartesian coordinates. This imposes severe restrictions on the potential $V(x,y)$ and the function $g(x,y)$, namely \cite{frivs1965higher}
\begin{equation}
V(x,y)=V_1(x)+V_2(y), \label{Potential-separabale}
\end{equation} 
\begin{equation}
g(x,y)=V_1(x)-V_2(y).
\end{equation}
The operators $X$ and $Y$ do not commute but generate a non-Abelian polynomial algebra.The Hamiltonian 	system \eqref{Hamiltonian},\eqref{PW-integral5-I},\eqref{integral5-I} is superintegrable, since it has more integrals of motion($n=3$) than degrees of freedom($n=2$). For precise definitions, properties of superintegrable systems and reasons why they are of physical and mathematical interest, see the review article \cite{miller2013classical}.\\
In writing the integral $X$ in \eqref{integral5-I} we use the general formalism introduced in Ref.~\onlinecite{post2015general} for $N$-th order quantum integrals in $E_2$.\\
The concept of ``exotic superintegrable systems'' was introduced and studied in several recent articles \cite{gravel2002superintegrability,gravel2004hamiltonians,tremblay2010third,marquette2007polynomial,marquette2009superintegrability,marquette2016connection,marquette2017fourth,escobar2017fourth}. These are superintegrable systems with $H$ as in \eqref{Hamiltonian}, $X$ as in  \eqref{integral5-I} separating in Cartesian or polar coordinates. They are ``exotic'' because of an additional requirement, namely that \textit{the potential $V(x,y)$ should not be the solution of any linear differential equation}. For potentials of the form \eqref{Potential-separabale} this implies that $V_1(x)$, $V_2(y)$ or possibly both $V_1(x)$ and $V_2(y)$  satisfy only nonlinear ODEs. It was observed in Ref.~\onlinecite{gravel2002superintegrability,gravel2004hamiltonians,marquette2009superintegrability,marquette2017fourth} that for $N=3$ and $N=4$ all of these nonlinear equations have the Painlev\'e property. This means that their solutions have no movable singularities other than poles \cite{ince1956ordinary,conte2008painleve,conte1999painleve,cosgrove2000chazy,cosgrove2000higher,bureau1971equations}.\\
The results for $N=3$ and $N=4$ suggest the following conjecture: superintegrable potentials in quantum mechanics that allow the separation of variables in the Schr\"{o}dinger equation in Cartesian or polar coordinates in $E_2$ satisfy ODEs that have the Painlev\'e property.\\
The purpose of the present article is to verify this ``Painlev\'e conjecture'' for $N=5$. In the process we observed that for $N=5$ the potential can be ``doubly exotic''. By this we mean that \textit{both $V_1(x)$ and $V_2(y)$ satisfy nonlinear equations with the Painlev\'e property}. This was also the case\cite{gravel2002superintegrability,gravel2004hamiltonians} for $N=3$, not however for $N=4$, where either $V_1(x)$ or $V_2(y)$ had to satisfy a linear equation. In this article we concentrate on doubly exotic potentials. ``Singly exotic'' ones are left for a future study. \\
The structure of the article is the following. In section $2$ we derive the determining equations governing the existence and form of the $N=5$ integral $X$ of eq.\eqref{integral5-I}. The expression for the commutator $[\mathcal{H},X]$ is a $6$-th order operator of the form 
	\begin{equation}
[\mathcal{H},X]=\sum\limits_{0 \leq{i+j}\leq6}M_{i,j}(x,y)\partial_{x_1}^i \partial_{x_2}^j=0. \label{commutator2}
\end{equation}
All coefficients $M_{i,j}$ must vanish. Some of the obtained determining equations depend explicitly on the Planck constant $\hbar$. The classical determining equations are obtained for $\hbar\rightarrow0$. We obtain a linear PDE as the compatibility condition for the $V(x,y)$. For exotic potentials this equation must be satisfied identically. In Section $3$ we restrict to separable potentials of the form \eqref{Potential-separabale} and obtain linear ODEs for $V_1(x)$ and $V_2(y)$. In Section $4$ we restrict further, namely to doubly exotic potentials and obtain an expression for the general integral of motion $X$ in terms of the potentials $V_1(x)$ and $V_2(y)$ and of some constants to be specified later. In section $5$ we solve the nonlinear ODEs for the potentials $V_1(x)$ and $V_2(y)$. The results are summed up and analyzed in Section $6$. Section $7$ is devoted to conclusions and future outlook.  
	\section{Conditions for the existence of a fifth order integral in quantum mechanics }  
  \ifx Equation (\ref{integral5-I}) defines a fifth-order Hermitian operator. In order to be an integral of motion, $X$ has to commute with the Hamiltonian (\ref{Hamiltonian}). Due to the fact that we have a real Hamiltonian and purely imaginary momentum operators, the terms of even order must commute independently of the terms of odd order. Thus, we can restrict ourselves to fifth-order integrals of the form: 
	\begin{equation}
	X=\frac{1}{2}\sum \limits_{l=0}^2\sum \limits_{j=0}^{5-2l}\{f_{j,2l}, p_1^j p_2^{5-2l-j}\}	
	\end{equation}
	The condition
	\begin{equation}
	[\mathcal H,X]=0
	\end{equation}
	is equivalent to an equation of the form:
	\begin{equation}
	\sum\limits_{0 \leq{i+j}\leq6}M_{i,j}(x,y)p_1^i p_2^j=0.
	\end{equation}
	\fi

	Setting $M_{i,j}(x,y)=0$, we find a set of $28$ differential equations, $12$ of which are consequences of the other $16$. More specifically, it is sufficient to set $M_{i,j}(x,y)=0$ with $i+j=2k, k=0,1,2,3$, in order to satisfy \eqref{commutator2} . Alternatively, we can use Theorem $2$ in Ref.~\onlinecite{post2015general} which gives directly the determining equations for the coefficients $f_{j,2l}$ in \eqref{integral5-I}. The first seven equations are given by setting $M_{i,j}=0$ with $i+j=6$. This gives the well known \textit{Killing-equations}:
	\begin{equation}
	{\frac{\partial f_{j-1,0} }{\partial x}}+{\frac{\partial f_{j,0} }{\partial y}} =0, \ \ 0\leq j\leq6 \label{Killing} ,
	\end{equation}
	with $f_{-1,0}=f_{6,0}=0$.
	Equations (\ref{Killing}) are the conditions for the highest order terms of $X$ to commute with the free Hamiltonian $\mathcal{H}_0=p_1^2+p_2^2$. They can be solved directly to give:
	\begin{equation}
	f_{j,0}=\sum _{n=0}^{5-j}\sum _{m=0}^j \binom{5-n-m}{j-m}A_{5-n-m,m,n} x^{5-j-n} (-y)^{j-m}. \label{f_j0}
	\end{equation}
	That is:
	\begin{eqnarray}
	&f_{50}=&A_{050}-y A_{140}+y^2 A_{230}-y^3 A_{320}+y^4 A_{410}-y^5 A_{500},\\
	&f_{40}=&A_{041}-y A_{131}+x A_{140}+y^2 A_{221}-2 x y A_{230}-y^3 A_{311}+3 x y^2 A_{320}+y^4A_{401}\nonumber \\& &  
	-4 x y^3 A_{410} +5 x y^4 A_{500}, \\
	&f_{30}=&A_{032}-y A_{122}+x A_{131}+y^2 A_{212}-2 x y A_{221}+x^2 A_{230}-y^3 A_{302}+3 x y^2
	A_{311}\nonumber\\
	& &-3 x^2 y A_{320}-4 x y^3 A_{401}+6 x^2 y^2 A_{410}-10 x^2 y^3 A_{500} ,\\
	&f_{20}=&A_{023}-y A_{113}+x A_{122}+y^2 A_{203}-2 x y A_{212}+x^2 A_{221}+3 x y^2 A_{302}-3 x^2 yA_{311}\nonumber\\
	& & +x^3 A_{320}+6 x^2 y^2 A_{401}-4 x^3 y A_{410}+10 x^3 y^2 A_{500} ,\\
	&f_{10}=&A_{014}-y A_{104}+x A_{113}-2 x y A_{203}+x^2 A_{212}-3 x^2 y A_{302}+x^3 A_{311}-4 x^3 y A_{401}\nonumber\\
	& &+x^4 A_{410}-5 x^4 y A_{500} ,\\
	&f_{00}=&A_{005}+x A_{104}+x^2 A_{203}+x^3 A_{302}+x^4 A_{401}+x^5 A_{500} .
	\end{eqnarray}
	where $A_{ijk}$ are constants.\\
	\begin{remark} At this point, the specific form of the functions $f_{j,0}$ allows us to rewrite the integral (\ref{integral5-I}) in a different symmetrized form, namely:
		\begin{eqnarray}
		X=& \frac{1}{2} \sum \limits_{m=0}^5 \sum \limits_{n=0}^{5-m} A_{5-m-n,m,n} \left\{L_3^{5-m-n},p_1^m
		p_2^n\right\}+ \frac{1}{2}\sum \limits_{l=1}^2\sum \limits_{j=0}^{5-2l}\{g_{j,2l}, p_1^j p_2^{5-2l-j}\} ,\nonumber\\ 
		& L= xp_2-yp_1.& \label{integral5-III}
		\end{eqnarray}
		The form in \eqref{integral5-III} introduced in Ref.~\onlinecite{post2015general} for arbitrary $N$ was the starting point in previous articles (for $2 \leqslant N \leqslant 4$)\cite{makarov1967systematic,gravel2004hamiltonians,tremblay2010third,marquette2017fourth}. It is important to notice that this choice of symmetrization affects the form of the functions $f_{j,2l}$ with $j\neq0$ in the original form of the integral. Thus, the functions $g_{j,2l}$ with $j\neq0$ must satisfy a different, though equivalent, set of differential equations in order to satisfy the commutation relation (\ref{commutator2}). Despite the fact that separable third and fourth order superintegrable systems have been studied using a form analogous to (\ref{integral5-III}), we will continue here assuming that the integral has the form in (\ref{integral5-I}) (with $N=5$) in order to use and verify the general results obtained in Ref.~\onlinecite{post2015general}.
	\end{remark}
	The next five independent determining equations are given by setting $M_{i,j}(x,y)=0$ with $i+j=4$. After some simplifications, these are (we use the notation $f_{j,k}{}^{(n,m)}:=  \partial_x^n \partial_y^m f_{j,k} $):
	\begin{align}
	f_{02}{}^{(0,1)}=&\frac{5}{2} f_{00} V_y+\frac{1}{2} f_{10} V_x +\hbar^2 \left(\frac{-3}{2} A_{311}+6 y A_{401}-6 x A_{410}+30 x yA_{500}\right), \label{D1}\\
	f_{12}{}^{(0,1)} + f_{02}{}^{(1,0)} =&2 f_{10} V_y+f_{20}V_x +\hbar^2 \left(3 A_{302}-3 A_{320}+12 x A_{401}+12 y A_{410} \nonumber \right.\\ &+ \left.
	30 x^2A_{500}-30 y^2 A_{500}\right) \label{D2} ,\\ 
	f_{22}{}^{(0,1)} +f_{12}{}^{(1,0)} =&\frac{3}{2} f_{20}
	V_y+\frac{3}{2} f_{30}V_x \label{D3} ,\\
	f_{32}{}^{(0,1)} +f_{22}{}^{(1,0)}=&f_{30} V_y+2 f_{40} V_x+\hbar^2 \left(3 A_{302}-3 A_{320}+12 x A_{401}+12 y A_{410}\nonumber \right.\\ &+ \left.
	30 x^2A_{500}-30 y^2 A_{500}\right) \label{D4}, \\
	f_{32}{}^{(1,0)} =&\frac{1}{2} f_{40}V_y+\frac{5}{2} f_{50}V_x+\hbar^2 \left(\frac{3}{2} A_{311}-6 y A_{401}+6 x A_{410}-30 x yA_{500}\right) \label{D5} .
	\end{align}
	Eq.\eqref{D1}- \eqref{D5}  are linear PDEs for $f_{a,2}$ and $V$ since $f_{j,0}$ ($j=0,1,...5$) are known from \eqref{f_j0}. We shall use their compatibility to obtain a linear partial differential equation satisfied by the potential $V(x,y)$. Next, setting $M_{i,j}(x,y)=0$, with $i+j=2$, gives us three more determining equations. After some simplifications using \eqref{D1}-\eqref{D5}, these are

	\begin{align}
	(f_{04}){}^{(0,1)} =&\frac{3}{2} f_{02}V_y+\frac{1}{2} f_{12}
	V_x+\hbar^2 \left[  \left(\frac{5}{8} f_{00}-\frac{1}{8}f_{40}\right) V_{yyy}
	+\left(\frac{15}{4}f_{00}{}-\frac{1}{4}f_{40}+f_{10}{}\nonumber \right.\right.\\
	 &+\left. \left.
	\frac{1}{4}f_{30}{}^{(1,0)}\right)V_{yy} 
	+\left(\frac{7}{4} f_{10}{}^{(0,1)}+\frac{1}{4}
	f_{30}{}^{(0,1)}-\frac{5}{4} f_{50}{}^{(0,1)}-\frac{5}{2}
	f_{00}{}^{(1,0)}\nonumber \right.\right.\\&+ \left.\left.
	\frac{1}{2} f_{20}{}^{(1,0)} 
	 +\frac{1}{2} f_{40}{}^{(1,0)}\right) V_{xy}
	+\left(\frac{5}{8}f_{10}+\frac{1}{4} f_{30}-\frac{5}{8} f_{50}\right)
	V_{xyy}\nonumber \right.\\&+ \left.
	\left(\frac{1}{2} f_{20}{}^{(0,1)}+\frac{1}{2}
	f_{40}{}^{(0,1)} 
	 -\frac{1}{2} f_{10}{}^{(1,0)}\right)V_{xx}+\left(\frac{-15}{8}f_{00}{}^{(0,2)}-\frac{1}{8} f_{40}{}^{(0,2)}\nonumber \right.\right.\\&- \left.\left.
	 \frac{1}{2}f_{10}{}^{(1,1)}+\frac{1}{4} f_{30}{}^{(1,1)}-\frac{5}{4}
	f_{00}{}^{(2,0)} -\frac{3}{8} f_{20}{}^{(2,0)}\right)V_{y} +\left(\frac{-3}{8}f_{10}{}^{(0,2)}\nonumber \right.\right.\\&- \left.\left.
	 \frac{5}{8} f_{50}{}^{(0,2)}-\frac{1}{4} f_{20}{}^{(1,1)}+\frac{1}{2}
	f_{40}{}^{(1,1)}-\frac{1}{4} f_{10}{}^{(2,0)} 
	-\frac{3}{8} f_{30}{}^{(2,0)}\right)V_x  \nonumber \right.\\&+ \left.
	\left(\frac{-5}{4}f_{00}+\frac{1}{8}f_{20}+\frac{1}{2} f_{40}\right) V_{xxy} 
	+\left(-\frac{1}{4}
	f_{10}-\frac{1}{8} f_{30}\right) V_{xxx} \right]  \label{D6},\\	
	f_{14}{}^{(0,1)}+f_{04}{}^{(1,0)}=&f_{12} V_y+f_{22}V_x+
	\hbar^2  \left[\left(\frac{5}{4}
	f_{00}+\frac{1}{4} f_{20}+\frac{1}{4} f_{40}\right)
	V_{xyy} +\left(\frac{1}{4}
	f_{10}+\frac{1}{4} f_{30} \nonumber \right.\right.\\&+ \left.\left.
	\frac{5}{4} f_{50}\right)V_{xxy}+ \left(f_{10}{}^{(0,1)}+\frac{5}{4}
	f_{00}{}^{(1,0)}+\frac{3}{4} f_{20}{}^{(1,0)}+\frac{1}{4}
	f_{40}{}^{(1,0)}\right)V_{yy}\nonumber \right.\\&+ \left.
	\left(\frac{1}{4} f_{10}{}^{(0,1)}+\frac{3}{4}
	f_{30}{}^{(0,1)}
	+\frac{5}{4} f_{50}{}^{(0,1)}+f_{40}{}^{(1,0)}\right) V_{xx}
	+\left(\frac{5}{4} f_{00}{}^{(0,1)}\nonumber \right.\right.\\&+ \left.\left.
	\frac{5}{4}
	f_{20}{}^{(0,1)}
	+\frac{1}{4} f_{40}{}^{(0,1)}+\frac{1}{4}
	f_{10}{}^{(1,0)}+\frac{5}{4} f_{30}{}^{(1,0)}
	+\frac{5}{4} f_{50}{}^{(1,0)}\right) V_{xy}\nonumber \right.\\&+ \left.
	\left(-f_{10}{}^{(0,2)}+\frac{5}{4} f_{00}{}^{(1,1)} 
	-\frac{3}{4}
	f_{20}{}^{(1,1)}+\frac{1}{4} f_{40}{}^{(1,1)}-\frac{1}{2}
	f_{30}{}^{(2,0)}\right)V_{y}\nonumber \right.\\&+ \left.
	\left(-\frac{1}{2}
	f_{20}{}^{(0,2)}+\frac{1}{4} f_{10}{}^{(1,1)}-\frac{3}{4}
	f_{30}{}^{(1,1)}
	+\frac{5}{4} f_{50}{}^{(1,1)}-f_{40}{}^{(2,0)}\right)V_{x}\right] \label{D7} ,\\
	f_{14}{}^{(1,0)}=&\frac{1}{2} f_{22}V_{y}+\frac{3}{2} f_{32}
	V_{x}+\hbar^2 \left[ \left(-\frac{1}{8} f_{20}-\frac{1}{4}
	f_{40}\right) V_{yyy}+\left(-\frac{1}{2}
	f_{40}{}^{(0,1)}\nonumber \right.\right.\\&+ \left.\left.
	\frac{1}{2} f_{10}{}^{(1,0)} +\frac{1}{2}
	f_{30}{}^{(1,0)}\right)V_{yy} 
	+\left(\frac{1}{2} f_{10}{}^{(0,1)}+\frac{1}{2}
	f_{30}{}^{(0,1)}-\frac{5}{2} f_{50}{}^{(0,1)}\nonumber \right.\right.\\&- \left.\left.
	\frac{5}{4}
	f_{00}{}^{(1,0)}+\frac{1}{4} f_{20}{}^{(1,0)}
	+\frac{7}{4}
	f_{40}{}^{(1,0)}\right) V_{xy}
	+\left(\frac{1}{2}
	f_{10}+\frac{1}{8} f_{30}-\frac{5}{4} f_{50}\right)
	V_{xyy}\nonumber \right.\\&+ \left.
	\left(\frac{1}{4}
	f_{20}{}^{(0,1)}+f_{40}{}^{(0,1)} 
	-\frac{1}{4}
	f_{10}{}^{(1,0)}+\frac{15}{4} f_{50}{}^{(1,0)}\right)
	V_{xx}
	+\left(\frac{-3}{8}
	f_{20}{}^{(0,2)}\nonumber \right.\right.\\&- \left.\left.
	\frac{1}{4} f_{40}{}^{(0,2)}+\frac{1}{2} f_{10}{}^{(1,1)} 
	-\frac{1}{4} f_{30}{}^{(1,1)}-\frac{5}{8}
	f_{00}{}^{(2,0)}-\frac{3}{8} f_{40}{}^{(2,0)}\right)V_{y}\nonumber \right.\\&+ \left.
	\left(\frac{-3}{8}f_{30}{}^{(0,2)}
	-\frac{5}{4} f_{50}{}^{(0,2)}+\frac{1}{4} f_{20}{}^{(1,1)}  
	-\frac{1}{2}
	f_{40}{}^{(1,1)}-\frac{1}{8} f_{10}{}^{(2,0)}\nonumber \right.\right.\\&- \left.\left.
	\frac{15}{8}
	f_{50}{}^{(2,0)}\right)V_{x}
	+\left(\frac{-5}{8} f_{00}+\frac{1}{4}
	f_{20}+\frac{5}{8} f_{40}\right) V_{xxy}
	+\left(-\frac{1}{8}
	f_{10}+\frac{5}{8} f_{50}\right) V_{xxx}. \right] \label{D8}.
	\end{align}
	\noindent
	The compatibility of these equations can be used to obtain a \textit{nonlinear} partial differential equation satisfied by the potential $V(x,y)$ and the undetermined coefficients $f_{j,2}$.\\ 
	Finally, setting $M_{0,0}(x,y)=0$ gives us the last determining equation relating all functions $f_{i,j}$ and the potential. After some simplifications using (\ref{D1})-(\ref{D5}), this equation reads
	
	\begin{align}
	0=&2 f_{04}V_{y}+2 f_{14}V_{x}+\hbar^2 \left[-\frac{1}{2} f_{02} V_{yyy}-\frac{1}{2} f_{12}V_{xyy}+V_{y} \left(\frac{-3}{2}f_{02}{}^{(0,2)}-f_{12}{}^{(1,1)}-\frac{1}{2}
	f_{22}{}^{(2,0)}\right)\nonumber \right.\\ &+\left. 
	V_{x} \left(-\frac{1}{2} f_{12}{}^{(0,2)}-f_{22}{}^{(1,1)}-\frac{3}{2}
	f_{32}{}^{(2,0)}\right)-\frac{1}{2} f_{22}V_{xxy}-\frac{1}{2} f_{32}V_{xxx}\right]
	+\hbar^4 \left[ \left(\frac{1}{8}f_{00}\nonumber \right. \right. \\ &- \left. \left.  
	\frac{1}{8} f_{40}\right) V_{yyyyy}
	+\left(-\frac{1}{8} f_{10}+\frac{1}{8} f_{50}\right) V_{xxxxx}+\left(\frac{1}{8} f_{10}+\frac{1}{4} f_{30}-\frac{5}{8}
	f_{50}\right) V_{xyyyy}\nonumber \right.\\ &+ \left.
	\left(\frac{-5}{8} f_{00}
	+\frac{1}{4} f_{20}
	+\frac{1}{8} f_{40}\right) V_{xxxxy}
	+\left(-\frac{1}{4} f_{20}+\frac{1}{2} f_{40}\right) V_{xxyyy}+\left(\frac{1}{2} f_{10}-\frac{1}{4} f_{30}\right)
	V_{xxxyy}\nonumber \right.\\ &+ \left.
	\left(-\frac{1}{2} f_{40}{}^{(0,1)}
	+\frac{1}{4}f_{30}{}^{(1,0)}\right)V_{yyyy}  +\left(\frac{1}{4} f_{20}{}^{(0,1)}-\frac{1}{2} f_{10}{}^{(1,0)}\right) V_{xxxx}
	+\left(\frac{3}{4} f_{30}{}^{(0,1)}\nonumber \right. \right. \\ &- \left. \left. 
	\frac{5}{2} f_{50}{}^{(0,1)}-\frac{3}{4}
	f_{20}{}^{(1,0)}  
	+\frac{1}{2} f_{40}{}^{(1,0)}\right) V_{xyyy} +\left(\frac{1}{2} f_{10}{}^{(0,1)} -\frac{3}{4} f_{30}{}^{(0,1)}-\frac{5}{2}
	f_{00}{}^{(1,0)}\nonumber \right. \right. \\ &+ \left. \left. 
	\frac{3}{4} f_{20}{}^{(1,0)}\right) V_{xxxy}
	+\left(\frac{-3}{4}f_{20}{}^{(0,1)}
	+\frac{3}{2} f_{40}{}^{(0,1)}+\frac{3}{2} f_{10}{}^{(1,0)}-\frac{3}{4} f_{30}{}^{(1,0)}\right) V_{xxyy} \nonumber \right. \\ &+ \left. 
	\left(\frac{5}{4} f_{00}{}^{(0,2)}-\frac{3}{4} f_{40}{}^{(0,2)}+\frac{1}{2}
	f_{10}{}^{(1,1)}+\frac{3}{4} f_{30}{}^{(1,1)}
	 -\frac{1}{4} f_{20}{}^{(2,0)}\right)V_{yyy} +\left(-\frac{1}{4} f_{30}{}^{(0,2)}\nonumber \right. \right. \\ &+ \left. \left. 
	 \frac{3}{4} f_{20}{}^{(1,1)}+\frac{1}{2} f_{40}{}^{(1,1)}
	-\frac{3}{4} f_{10}{}^{(2,0)}+\frac{5}{4} f_{50}{}^{(2,0)}\right)
	V_{xxx} +\left(\frac{3}{4} f_{10}{}^{(0,2)}+\frac{3}{4} f_{30}{}^{(0,2)}\nonumber \right. \right. \\ &- \left. \left. 
	\frac{15}{4}
	f_{50}{}^{(0,2)}-\frac{3}{4} f_{20}{}^{(1,1)}+\frac{3}{2} f_{40}{}^{(1,1)}+\frac{3}{2} f_{10}{}^{(2,0)}\right)V_{xyy} 
	+\left(\frac{3}{2} f_{40}{}^{(0,2)}
	+\frac{3}{2} f_{10}{}^{(1,1)}\nonumber \right. \right. \\ &- \left. \left. 
	\frac{3}{4} f_{30}{}^{(1,1)}-\frac{15}{4} f_{00}{}^{(2,0)}+\frac{3}{4} f_{20}{}^{(2,0)}+\frac{3}{4} f_{40}{}^{(2,0)}\right) V_{xxy}
	+\left(\frac{1}{2} f_{40}{}^{(0,3)}-\frac{3}{4} f_{30}{}^{(1,2)}\nonumber  \right. \right. \\ &+\left. \left.  
	\frac{3}{4} f_{20}{}^{(2,1)}-\frac{1}{2} f_{10}{}^{(3,0)}\right)V_{xx}
	+ \left(-\frac{1}{2} f_{40}{}^{(0,3)}+\frac{3}{4}
	f_{30}{}^{(1,2)}-\frac{3}{4} f_{20}{}^{(2,1)}+\frac{1}{2} f_{10}{}^{(3,0)}\right)V_{yy}\nonumber \right.\\ &+ \left.
	\left(\frac{1}{4}
	f_{30}{}^{(0,3)}-\frac{5}{2} f_{50}{}^{(0,3)}-\frac{3}{4} f_{20}{}^{(1,2)}+\frac{3}{2} f_{40}{}^{(1,2)}+\frac{3}{2}
	f_{10}{}^{(2,1)}-\frac{3}{4} f_{30}{}^{(2,1)}\nonumber \right. \right. \\&- \left. \left. 
	\frac{5}{2} f_{00}{}^{(3,0)}(x,y)
	+\frac{1}{4} f_{20}{}^{(3,0)}\right)V_{xy} 
	+ \left(\frac{5}{8} f_{00}{}^{(0,4)}-\frac{1}{8} f_{40}{}^{(0,4)}+\frac{1}{2} f_{10}{}^{(1,3)}+\frac{1}{4} f_{30}{}^{(1,3)}\nonumber \right. \right. \\&+ \left. \left. 
	\frac{1}{2}
	f_{10}{}^{(3,1)}+\frac{1}{4} f_{30}{}^{(3,1)}
	-\frac{5}{8} f_{00}{}^{(4,0)}+\frac{1}{8}
	f_{40}{}^{(4,0)}\right)V_{y}
	+ \left(\frac{1}{8} f_{10}{}^{(0,4)}-\frac{5}{8} f_{50}{}^{(0,4)}+\frac{1}{4}
	f_{20}{}^{(1,3)} \nonumber \right. \right. \\ &+ \left. \left. 
	\frac{1}{2} f_{40}{}^{(1,3)}+\frac{1}{4} f_{20}{}^{(3,1)}+\frac{1}{2} f_{40}{}^{(3,1)}-\frac{1}{8}
	f_{10}{}^{(4,0)}+\frac{5}{8} f_{50}{}^{(4,0)}\right)V_{x} \right] \label{D9} .
	\end{align}
	\begin{remark}
		In their original form, equations (\ref{D6})-(\ref{D9}) were polynomials of degree six in $\hbar$. However, the terms proportional to $\hbar^6$ vanished because of the polynomial form (\ref{f_j0}) of the functions $f_{j,0}$. 
	\end{remark}
	As noted above, equations (\ref{D1})-(\ref{D5}) must be compatible, that is:
	\begin{eqnarray}
	&\partial_{xxxx}(f_{02}{}^{(0,1)})&-\partial_{xxxy}(f_{12}{}^{(0,1)} + f_{02}{}^{(1,0)})+\partial_{xxyy}(f_{22}{}^{(0,1)} + f_{12}{}^{(1,0)}) \label{compatibilité_linéaire1}\\ && \nonumber
	-\partial_{xyyy}(f_{32}{}^{(0,1)} + f_{22}{}^{(1,0)}) +\partial_{yyyy}(f_{32}{}^{(1,0)})=0. 
	\end{eqnarray}
	Substituting the expressions of $f_{j,2}$ in \eqref{compatibilité_linéaire1}, we obtain a fifth-order linear PDE for the potential $V(x,y)$, namely
	\begin{align}
	&\frac{1}{2} f_{40} V_{yyyyy}+\frac{1}{2} f_{10} V_{xxxxx}+\left(-f_{30}+\frac{5}{2}f_{50}\right) V_{xyyyy}+\left(\frac{5}{2} f_{00}-f_{20}\right) V_{xxxxy}+\Big(\frac{3}{2} f_{20}  \nonumber \\ &- 
	2f_{40}\Big)V_{xxyyy}
	+\left(-2 f_{10}+\frac{3}{2} f_{30}\right)
	V_{xxxyy}+\left(2f_{40}{}^{(0,1)}-f_{30}{}^{(1,0)}\right)V_{yyyy} 
	+\left(-f_{20}{}^{(0,1)}\nonumber \right. \\ &+  \left.
	2 f_{10}{}^{(1,0)}\right)V_{xxxx}
	+\left(-3f_{30}{}^{(0,1)}+10 f_{50}{}^{(0,1)}+3 f_{20}{}^{(1,0)}-2f_{40}{}^{(1,0)}\right) V_{xyyy}
	+\left(-2 f_{10}{}^{(0,1)}\nonumber \right. \\ &+  \left.
	3f_{30}{}^{(0,1)} +10 f_{00}{}^{(1,0)}-3 f_{20}{}^{(1,0)}\right)V_{xxxy}+\left(3 f_{40}{}^{(0,2)}-3f_{30}{}^{(1,1)} 
	+\frac{3}{2} f_{20}{}^{(2,0)}\right)V_{yyy} \nonumber\\ &+\left(\frac{3}{2} f_{30}{}^{(0,2)}-3f_{20}{}^{(1,1)} 
	+3 f_{10}{}^{(2,0)}\right) V_{xxx} + \Big(-3 f_{30}{}^{(0,2)}+15 f_{50}{}^{(0,2)}+6 f_{20}{}^{(1,1)}\nonumber \\ &-  \left.
	6f_{40}{}^{(1,1)}-6 f_{10}{}^{(2,0)}+\frac{3}{2}
	f_{30}{}^{(2,0)}\right)V_{xyy}
	+\left(\frac{3}{2} f_{20}{}^{(0,2)}-6
	f_{40}{}^{(0,2)}-6 f_{10}{}^{(1,1)}+6 f_{30}{}^{(1,1)}\nonumber \right. \\&+  \left. 
	15 f_{00}{}^{(2,0)}-3 f_{20}{}^{(2,0)}\right) V_{xxy}
	+\left(3 f_{20}{}^{(0,1)}
	-6 f_{40}{}^{(0,1)}
	-6 f_{10}{}^{(1,0)}+3 f_{30}{}^{(1,0)}\right) V_{xxyy}
\nonumber \\ &+\left(2 f_{40}{}^{(0,3)}-3 f_{30}{}^{(1,2)}+3 f_{20}{}^{(2,1)}-2 f_{10}{}^{(3,0)}\right)V_{yy} +\left(-2 f_{40}{}^{(0,3)}
	+3 f_{30}{}^{(1,2)}-3 f_{20}{}^{(2,1)}\nonumber \right. \\&+\left.
	2f_{10}{}^{(3,0)}\right)V_{xx} 
	+\left(-f_{30}{}^{(0,3)}+10
	f_{50}{}^{(0,3)}+3 f_{20}{}^{(1,2)}-6 f_{40}{}^{(1,2)}
	-6 f_{10}{}^{(2,1)}
	+3 f_{30}{}^{(2,1)}\nonumber \right. \\&+\left.
	10 f_{00}{}^{(3,0)}-f_{20}{}^{(3,0)}\right)V_{xy} + \left(\frac{1}{2}
	f_{40}{}^{(0,4)}-f_{30}{}^{(1,3)}+\frac{3}{2} f_{20}{}^{(2,2)}-2
	f_{10}{}^{(3,1)} +\frac{5}{2} f_{00}{}^{(4,0)}\right)V_{y}\nonumber \\&+ \left(\frac{5}{2} f_{50}{}^{(0,4)}-2 f_{40}{}^{(1,3)}+\frac{3}{2}
	f_{30}{}^{(2,2)}-f_{20}{}^{(3,1)}+\frac{1}{2}
	f_{10}{}^{(4,0)}\right)V_{x} =0 \label{compatibilté_linéaire2}.
	\end{align}
	Note that equation \eqref{compatibilté_linéaire2} does not depend on $\hbar$ and is thus valid both in classical and quantum mechanics. Indeed, this is true for an arbitrary $Nth$-order integral of motion \cite{post2015general} with $N\geq2$.\\
	Using \eqref{D6}-\eqref{D8} and their compatibility condition
	\begin{equation}
	\partial_{xx}(f_{04}{}^{(0,1)})-\partial_{xy}(f_{14}{}^{(0,1)} + f_{04}{}^{(1,0)})+\partial_{yy}(f_{14}{}^{(1,0)})=0 , \label{compatibilité_non_linéaire1}
	\end{equation}
	we find a fifth-order nonlinear PDE for $V(x,y)$ and $f_{j2}$ $(j=0,...,3)$ which \textit{does} depend on $\hbar$. Equations \eqref{D6}-\eqref{D8} involve $5$ unknown functions of $2$ variables namely $V(x,y)$ and $f_{a2}(x,y), a=0,1,2,3$. These equations are compatible under condition \eqref{compatibilité_non_linéaire1}. This condition, given expliciitly in Appendix \ref{appendix:graph}, is nonlinear and at this stage we do not use it. We list it in this article because it will be useful in any study of $5$th order integrability\\
	We now pass over to the question of superintegrability. Thus we can assume that we have two further second order integrals $\mathcal{H}_1$ and  $\mathcal{H}_2$ in \eqref{Super-Integrals} and that $V(x,y)$ separates as in \eqref{Potential-separabale}. The linear and nonlinear compatibility conditions are used (see e.g. \eqref{compatibilité_non_linéaire2} below) after the variables are separated.   
	\section{Potentials separable in Cartesian coordinates}
	\noindent
	Equations \eqref{D1}-\eqref{D9} together with \eqref{Killing} assure the integrablity of the system, with two integrals of motion \eqref{Hamiltonian} and \eqref{integral5-I}. In general, it is difficult to solve the PDEs we encountered previously. However, by assuming that the potential in the Hamiltonian \eqref{Hamiltonian} has the form \eqref{Potential-separabale},
	we can greatly simplify the determining equations. In addition, this assumption gives a sufficient condition for the system to be maximally superintegrable, for we have now three integrals of motion, namely
	\begin{equation}
	\mathcal{H}_1=p_1^2	+V_{1}(x), \ \ \mathcal{H}_2=p_2^2	+V_{2}(y), \ \ X=\frac{1}{2}\sum \limits_{l=0}^2\sum \limits_{j=0}^{5-2l}\{f_{j,2l}, p_1^j p_2^{5-2l-j}\} . \label{Super-Integrals}
	\end{equation}
	Substituting \eqref{Potential-separabale} and  the polynomial functions $f_{j,0}$ into the linear compatibility \eqref{compatibilté_linéaire2}, we obtain
	\begin{align}
	&(720 A_{410}-3600 y A_{500}) V_1'(x)+\left(240 A_{311}-960 y A_{401}+960 x A_{410}-4800 x y A_{500}\right) V_1''(x)\nonumber \\
	&+\left(60 A_{212}-180 y A_{302}+180 x A_{311}-720 x y
	A_{401}+360 x^2 A_{410}-1800 x^2 y A_{500}\right) V_1{}^{(3)}(x)\nonumber\\
	&+\left(12 A_{113}-24 y A_{203}+24 x
	A_{212}-72 x y A_{302}+36 x^2 A_{311}-144 x^2 y A_{401}+48 x^3 A_{410}\nonumber \right.\\ 
	&-\left.  240 x^3 y A_{500}\right)V_1{}^{(4)}(x)
	+\left(2 A_{014}-2 y A_{104}+2 x A_{113}-4 xy A_{203}+2 x^2 A_{212}-6 x^2 y A_{302}\nonumber \right. \\
	&+\left. 2 x^3 A_{311}-8 x^3 y A_{401}+2 x^4 A_{410}-10 x^4 y A_{500}\right) V_1{}^{(5)}(x)
	+\left(720 A_{401}+3600 x A_{500}\right) V_2'(y)\nonumber\\
	&+\left(-240 A_{311}+960 y A_{401}
	-960 x A_{410}+4800 x y A_{500}\right) V_2''(y)+\left(60 A_{221}-180 y A_{311}\nonumber\right. \\ 
	&+\left.180 x A_{320}+360 y^2 A_{401}-720 x y A_{410}
	+1800 xy^2 A_{500}\right) V_2{}^{(3)}(y)+\left(-12 A_{131}+24 y A_{221} \nonumber \right. \\ 
	&-\left.24 xA_{230}-36 y^2 A_{311}+72 x y A_{320}+48 y^3 A_{401} -144 x y^2 A_{410}+240 xy^3 A_{500}\right) V_2{}^{(4)}(y)\nonumber\\
	&+\left(2 A_{041}-2 y A_{131}+2 xA_{140}+2 y^2 A_{221}-4 x y A_{230}
	-2 y^3 A_{311}+6 x y^2 A_{320}+2 y^4
	A_{401}\nonumber \right. \\ 
	&- \left. 8 x y^3 A_{410}+10 x y^4 A_{500}\right) V_2{}^{(5)}(y)=0. \label{compatibilité_linéaire3}
	\end{align}
	Equation \eqref{compatibilité_linéaire3} amounts to a system of ODEs since they involve two functions of one variable each and the coefficients depend on the powers $x^ay^b$. We differentiate \eqref{compatibilité_linéaire3} twice with respect to $x$ and collect the terms involving the same powers of $y$. The resulting equation is a polynomial  of degree one in $y$ with coefficients that are functions of $x$. As each of these two coefficients must vanish, we obtain two linear seventh order ODEs for $V_1(x)$, namely 
	\begin{subequations}
		\begin{align}
		&3360 A_{410} V_1{}^{(3)}(x)+\left(672 A_{311}+2688 x A_{410}\right)
		V_1{}^{(4)}(x)+\left(112 A_{212}+336 x A_{311}\nonumber \right. \\ &+ \left.
		672 x^2 A_{410}\right) V_1{}^{(5)}(x)
		+\left(16 A_{113}+32 x A_{212}+48 x^2 A_{311}+64 x^3
		A_{410}\right) V_1{}^{(6)}(x)\nonumber \\
		&+\left(2 A_{014}+2 x A_{113}+2 x^2 A_{212}+2 x^3
		A_{311}+2 x^4 A_{410}\right) V_1{}^{(7)}(x) \label{linear-ODE1x}=0 ,\\
		\nonumber \\ 
		&-16800 A_{500} V_1{}^{(3)}(x)-\left(2688 A_{401}+13440 x A_{500}\right)
		V_1{}^{(4)}(x)+\left(-336 A_{302}-1344 x A_{401}\right. \nonumber \\ &
		-\left. 3360 x^2 A_{500}\right) V_1{}^{(5)}(x)
		+\left(-32 A_{203}-96 x A_{302}-192 x^2 A_{401}-320 x^3
		A_{500}\right) V_1{}^{(6)}(x)\nonumber \\
		&+\left(-2 A_{104}-4 x A_{203}-6 x^2 A_{302}-8 x^3A_{401}-10 x^4 A_{500}\right) V_1{}^{(7)}(x)=0	 \label{linear-ODE2x}.
		\end{align}
	\end{subequations}
	Similarly, differentiating \eqref{compatibilité_linéaire3} twice with respect to $y$ and collecting the terms involving the same powers of $x$, we obtain two seventh order ODEs for $V_2(y)$, namely
	\begin{subequations}
		\begin{align}
		&3360 A_{401} V_2{}^{(3)}(y)+\left(-672 A_{311}+2688 y A_{401}\right)
		V_2{}^{(4)}(y)+\left(112 A_{221}-336 y A_{311}\right. \nonumber \\ &+\left.672 y^2 A_{401}\right) V_2{}^{(5)}(y)
		+\left(-16 A_{131}+32 y A_{221}-48 y^2 A_{311}+64 y^3
		A_{401}\right) V_2{}^{(6)}(y)\nonumber \\
		&+\left(2 A_{041}-2 y A_{131}+2 y^2 A_{221}-2 y^3
		A_{311}+2 y^4 A_{401}\right) V_2{}^{(7)}(y)=0	\label{linear-ODE1y} ,\\
		\nonumber \\
		&16800 A_{500} V_2{}^{(3)}(y)+\left(-2688 A_{410}+13440 y A_{500}\right)
		V_2{}^{(4)}(y)+\left(336 A_{320}-1344 y A_{410}\right. \nonumber \\ &+\left.
		3360 y^2 A_{500}\right) V_2{}^{(5)}(y)
		+\left(-32 A_{230}+96 y A_{320}-192 y^2 A_{410}+320 y^3
		A_{500}\right) V_2{}^{(6)}(y)\nonumber \\
		&+\left(2 A_{140}-4 y A_{230}+6 y^2 A_{320}-8 y^3
		A_{410}+10 y^4 A_{500}\right) V_2{}^{(7)}(y)=0	\label{linear-ODE2y}.
		\end{align}
	\end{subequations}
	\section{Doubly exotic superintegrable potentials}
	\noindent
	In this study, our task is to find and classify all doubly exotic potentials separating in Cartesian coordinates, that is all separable potentials that \textit{do not} satisfy any linear ODE. Consequently, we will request that all linear ODEs satisfied by $V_1(x)$ or $V_2(y)$ must vanish trivially (i.e their coefficients must be set to $0$ ). Thus, all four linear equations \eqref{linear-ODE1x}-\eqref{linear-ODE2y} must be trivially satisfied, that is
	\begin{equation}
	A_{500}=A_{401}=A_{410}=A_{311}=A_{302}=A_{212}=A_{203}=A_{113}=A_{104}=A_{014}=0,
	\end{equation}
	\begin{equation}
	A_{500}=A_{401}=A_{410}=A_{311}=A_{320}=A_{221}	=A_{230}=A_{131}=A_{140}=A_{041}=0,
	\end{equation}
    for \eqref{linear-ODE1x},\eqref{linear-ODE2x} and \eqref{linear-ODE1y},\eqref{linear-ODE2y} respectively.\\   	
	With these constraints, the integral of motion $X$ reads
	\begin{align}
	X=&\frac{1}{2} \left(\left\{p_1^5,A_{050}\right\}+\left\{p_1^3 p_2^2,A_{032}-y
	A_{122}\right\}+\left\{p_1^2 p_2^3,A_{023}+xA_{122}\right\}+\left\{p_2^5,A_{005}\right\}+\left\{p_1^3,f_{32}\right\}\right.\nonumber \\&+\left.
	\left\{p_2p_1^2,f_{22}\right\}+\left\{p_2^2p_1,f_{12}\right\}+\left\{p_2^3,f_{02}\right\}+\left\{p_1,f_{1,4}\right\}+\left\{p_2,f_{04}\right\}\right) , \label{integral5-IIII}
	\end{align}
	and the determining equations \eqref{D1}-\eqref{D9} take a much simpler form. For $f_{j2}$ we obtain
	\begin{align}
	f_{02}{}^{(0,1)}=&\frac{5}{2} A_{005} V_2'(y) , \label{D'1}\\
	f_{12}{}^{(0,1)}+f_{02}{}^{(1,0)}=&\left(A_{023}+x A_{122}\right) V_1'(x) ,\label{D'2}\\
	f_{22}{}^{(0,1)}+f_{12}{}^{(1,0)}=&\frac{3}{2} \left(A_{032}-y A_{122}\right) V_1'(x)+\frac{3}{2}  \left(A_{023}+x A_{122}\right) \label{D'3}
	V_2'(y) ,\\
	f_{32}{}^{(0,1)}+ f_{22}{}^{(1,0)}=&\left(A_{032}-y A_{122}\right) V_2'(y) ,\label{D'4}\\
	f_{32}{}^{(1,0)}=&\frac{5}{2} A_{050} V_1'(x)\label{D'5}.
	\end{align}
For $f_{j4}$ the determining equations reduce to 	
		\begin{align}
	f_{04}{}^{(0,1)}=&\frac{1}{2} f_{12} V_1'(x)+\frac{3}{2} f_{02}V_2'(y)\nonumber\\&
	+\frac{\hbar^2}{8} \left(5A_{005} V_2{}^{(3)}(y) -\left(A_{032}-y A_{122}\right)V_1{}^{(3)}(x)\right) , \label{D'6}\\
	f_{14}{}^{(0,1)}+f_{04}{}^{(1,0)}=&f_{22}V_1'(x)+f_{12}
	V_2'(y)+\frac{3\hbar^2A_{122} }{4} \left( -V_1''(x)+V_2''(y)\right) ,	\label{D'7}\\
	f_{14}{}^{(1,0)}=&\frac{1}{2} f_{22}V_2'(y)+\frac{3}{2} f_{32} V_1'(x)\nonumber\\&
	+\frac{ \hbar^2}{8} \left(5 A_{050} V_1{}^{(3)}(x)-\left(A_{023}+x A_{122}\right) V_2{}^{(3)}(y)\right) ,
	\label{D'8}\\
	0=&2 f_{14} V_1'(x)+2 f_{04} V_2'(y)-\frac{\hbar^2}{2} \left(f_{32}
	V_1{}^{(3)}(x)+f_{02} V_2{}^{(3)}(y)+3 V_2'(y) f_{02}{}^{(0,2)}\right. \nonumber\\&+ \left.
	V_1'(x)f_{12}{}^{(0,2)}
	+2 V_2'(y) f_{12}{}^{(1,1)}+2 V_1'(x)
	f_{22}{}^{(1,1)}+V_2'(y) f_{22}{}^{(2,0)}\right. \nonumber\\&+ \left.
	3V_1'(x)f_{32}{}^{(2,0)}\right) 
	+\frac{\hbar^4}{8} \left(A_{050}
	V_1{}^{(5)}(x)+A_{005} V_2{}^{(5)}(y)\right).\label{D'9}
\end{align}
	If we integrate directly \eqref{D'1}-\eqref{D'3} and \eqref{D'5}, we find
	\begin{eqnarray}
	f_{02}&=&\alpha _1(x)+\frac{5}{2} A_{005} V_2(y) ,	\label{f_02}\\
	f_{12}&=&\alpha _2(x)+y \left(A_{023} V_1'(x)+x A_{122} V_1'(x)-\alpha
	_1'(x)\right) ,	\label{f_12}\\
	f_{22}&=&\alpha _3(x)+\frac{3}{2} A_{023} V_2(y)+\frac{3}{2} x A_{122}
	V_2(y)+\frac{3}{2} y A_{032} V_1'(x)-\frac{5}{4} y^2 A_{122} V_1'(x) -y \alpha
	_2'(x)\nonumber \\
	& & -\frac{1}{2} y^2 A_{023} V_1''(x)-\frac{1}{2} x y^2 A_{122}
	V_1''(x)+\frac{1}{2} y^2 \alpha _1''(x) ,\\
	f_{32}&=&\alpha _4(y)+\frac{5}{2} A_{050} V_1(x), 	\label{f_32}
	\end{eqnarray}
	where $\alpha_i(x), i=1,2,3$ and $\alpha_4(y)$ are four unknown functions.\\
	When we substitute \eqref{f_02}-\eqref{f_32} into \eqref{D'4}, we find
	\begin{align}
	&-3 A_{122} V_2(y)+2 A_{032} V_2'(y)-2 y A_{122} V_2'(y)-2 \alpha _3'(x)-2 \alpha
	_4'(y)-3 y A_{032} V_1''(x)+\frac{7}{2} y^2 A_{122} V_1''(x)\nonumber \\
	&+2 y \alpha
	_2''(x)+y^2 A_{023} V_1{}^{(3)}(x)+x y^2 A_{122} V_1{}^{(3)}(x)-y^2 \alpha
	_1{}^{(3)}(x)=0. \label{compa_du milieu 1}	
	\end{align}
	We differentiate \eqref{compa_du milieu 1} twice with respect to $y$  and once with respect to $x$. When we integrate the resulting equation, we obtain
	\begin{equation}
	\alpha _1(x)=C_1+x C_2+x^2 C_3+x^3 C_4+\frac{1}{2} W_1(x) A_{122}+A_{023} W_1(x)'+x A_{122} W_1(x)',
	\end{equation}
	where $C_i$ are arbitrary constants of integration and $W_1(x)$,$W_2(y)$ are two auxiliary functions verifying
	\begin{equation}
	W_1'(x)=V_1(x), \ \ W_2'(y)=V_2(y). \label{W1&W2}
	\end{equation}
	Proceeding in this way for the other mixed derivatives of \eqref{compa_du milieu 1}, we find 
	\begin{subequations}
		\begin{eqnarray}
		\alpha _2(x)&=&C_5+x C_6+x^2 C_7+\frac{3}{2} A_{032} V_1(x),\\
		\alpha _3(x)&=&C_8+x C_9 ,\\
		\alpha _4(y)&=&C_{10}-y^3 C_4+y^2 C_7-y C_9-\frac{1}{2} W_2(y) A_{122}+\left(A_{032}-y A_{122}\right) W_2'(y).
		\end{eqnarray}
	\end{subequations}
	Thus the coefficients $f_{j2}$ are known in terms of $W_1(x)$,$W_2(y)$ and the constants $C_1, C_2,..., C_{10}$.  Next, we integrate \eqref{D'6} and \eqref{D'8} in order to determine $f_{j4}$ $(j=0,1)$ :
	\begin{align}
	f_{04}=&\frac{15}{8} A_{005} W_2(y)'{}^2+W_2'(y) \left(\frac{3 C_1}{2}+\frac{3 x C_2}{2}+\frac{3 x^2
	C_3}{2}+\frac{3 x^3 C_4}{2}+\left(\frac{3}{2} A_{023}+\frac{3}{2} x A_{122}\right) W_1'(x)\right.\nonumber\\ &+\left.\frac{3}{4} A_{122}
W_1(x)\right)+\beta _1(x)+\left(-\frac{1}{4} y^2 C_2-\frac{1}{2} x y^2 C_3-\frac{3}{4} x^2 y^2 C_4+\frac{y
	C_5}{2}+\frac{1}{2} x y C_6\right.\nonumber\\ &+\left.
\frac{1}{2} x^2 y C_7\right)  W_1''(x) +\left.\left(\frac{3}{4} y A_{032}-\frac{3}{8} y^2
A_{122}\right) W_1(x)' W_1''(x)+\frac{5}{8} \hbar^2 A_{005} W_2^{(3)}(y) \right.\nonumber\\ &+\left. 
\left(-\frac{1}{8} \hbar^2 y A_{032}+\frac{1}{16}
\hbar^2 y^2 A_{122}\right) W_1^{(4)}(x), \label{f_04}\right.\\
f_{14}=&\frac{15}{8} A_{050} W_1'(x){}^2+W_1'(x) \left(\frac{-3y^3 C_4}{2}+\frac{3 y^2 C_7}{2}-\frac{3 y
		C_9}{2}+\frac{3 C_{10}}{2}+\left(\frac{3}{2} A_{032}-\frac{3}{2} y A_{122}\right) W_2'(y)\right.\nonumber \\ &-\left. \frac{3}{4} A_{122}
	W_2(y)\right)+\beta _2(y)+\left(\frac{1}{2} x y^2 C_3+\frac{3}{4} x^2 y^2 C_4-\frac{1}{2} x y C_6-\frac{1}{2} x^2 y
	C_7+\frac{x C_8}{2}\right.\nonumber\\ &+\left.
	+\frac{x^2 C_9}{4}\right) W_2''(y)\left.+\left(\frac{3}{4} x A_{023}+\frac{3}{8} x^2 A_{122}\right)
	W_2'(y) W_2''(y)+\frac{5}{8} \hbar^2 A_{050} W_1^{(3)}(x)\nonumber \right.\\ & \left.
	+\left(-\frac{1}{8} \hbar^2 x A_{023}-\frac{1}{16} \hbar^2 x^2
	A_{122}\right) W_2{}^{(4)}(y), \label{f_14}\right.
	\end{align}
	where $\beta _1(x)$ and $\beta _2(y)$ are two unknown functions. They must be determined by substituting $f_{04}$ and $f_{14}$ into the remaining determining equation \eqref{D'7}. \\
	The nonlinear compatibility \eqref{compatibilité_non_linéaire1} of \eqref{D'6}-\eqref{D'8} gives
	\begin{align}
	&\left(-18 y C_4+6 C_7\right) V_1'(x)+\left(6 C_3+18 x C_4\right) V_2'(y)+\left(-4 y C_3-12 x y C_4+2 C_6+4 x
C_7\right) V_1''(x)\nonumber \\
	&+\left(4 y C_3+12 x y C_4-2 C_6-4 x C_7\right) V_2''(y)+\left(\frac{9}{4} A_{032}-\frac{9}{4}
	y A_{122}\right) V_1'(x) V_1''(x)
	+\left(\frac{9}{4} A_{023}\right. \nonumber\\ &+\left.
	\frac{9}{4} x A_{122}\right) V_2'(y)V_2''(y)+\left(-\frac{1}{2} \left(y C_2\right)-x y C_3-\frac{3}{2} x^2 y C_4+\frac{C_5}{2}+\frac{x
	C_6}{2}+\frac{x^2 C_7}{2}\right) V_1{}^{(3)}(x) \nonumber \\
	&+\left(\frac{y^2 C_3}{2}+\frac{3}{2} x y^2 C_4-\frac{y C_6}{2}-x y C_7
	+\frac{C_8}{2}+\frac{x C_9}{2}\right) V_2{}^{(3)}(y)+\left(\frac{3}{4} A_{032}-\frac{3}{4} y A_{122}\right)
	V_1(x) V_1{}^{(3)}(x)\nonumber \\ 
	&+\left(\frac{3}{4} A_{023}+\frac{3}{4} x A_{122}\right)V_2(y) V_2{}^{(3)}(y)
	+\left(-\frac{1}{8} \hbar^2 A_{032}+\frac{1}{8} \hbar^2 y A_{122}\right)
	V_1{}^{(5)}(x)\nonumber \\
	&+\left(-\frac{1}{8} \hbar^2 A_{023}-\frac{1}{8} \hbar^2 x A_{122}\right) V_2{}^{(5)}(y)=0. \label{compatibilité_non_linéaire2}
	\end{align}
	As in \eqref{compatibilité_linéaire3}, we differentiate \eqref{compatibilité_non_linéaire2} twice with respect to $x$ and collect the terms involving the same powers of $y$. The resulting equation gives us two nonlinear ODEs for $V_1(x)$ that can be integrated and we obtain two fourth order ODEs: 
	\begin{align}
	\hat{K}_1+x \hat{K}_2+x^2 \hat{K}_3=&3 C_7 V_1(x)+\frac{3}{2} C_6 V_1'(x)+\frac{1}{2} x C_6 V_1''(x)+3 x C_7 V_1'(x)+\frac{3}{4} A_{032}V_1'(x){}^2\nonumber \\
	&+\frac{1}{2} C_5 V_1''(x)+\frac{1}{2} x^2 C_7 V_1''(x)+\frac{3}{4} A_{032} V_1(x) V_1''(x)-\frac{1}{8}
	\hbar^2 A_{032} V_1{}^{(4)}(x),  \label{1 nonlinearODE for V_1}\\ 
	K_1+x K_2 +x^{2}K_3 =&-9 C_4 V_1(x)-9 x C_4 V_1'(x)-3 C_3 V_1'(x)-\frac{3}{4} A_{122} V_1'(x){}^2-\frac{1}{2} C_2 V_1''(x)\nonumber \\
	&-x C_3
	V_1''(x)-\frac{3}{2} x^2 C_4 V_1''(x)-\frac{3}{4} A_{122} V_1(x) V_1''(x)+\frac{1}{8} \hbar^2 A_{122}
	V_1{}^{(4)}(x) , \label{2 nonlinearODE for V_1}  
	\end{align}
	where  $K_i $ and $\hat{K}_i$ are arbitrary constants of integration.\\
	Similarly, we obtain two nonlinear ODEs for $V_2(y)$:
	\begin{align}
	\hat{D}_1+y \hat{D}_2+y^2 \hat{D}_3=&3 C_3 V_2(y)+3 y C_3 V_2'(y)-\frac{3}{2} C_6 V_2'(y)+\frac{3}{4} A_{023} V_2'(y){}^2+\frac{1}{2} y^2 C_3
	V_2''(y)\nonumber \\
	&-\frac{1}{2} y C_6 V_2''(y)+\frac{1}{2} C_8 V_2''(y)+\frac{3}{4} A_{023} V_2(y) V_2''(y)-\frac{1}{8} \hbar^2
	A_{023} V_2{}^{(4)}(y) ,\label{1 nonlinearODE for V_2}\\
	D_1+y D_2 +y^2 D_3=&9 C_4 V_2(y)+9 y C_4 V_2'(y)-3 C_7 V_2'(y)+\frac{3}{4} A_{122} V_2'(y){}^2+\frac{3}{2} y^2 C_4 V_2''(y)\nonumber \\&
	-y C_7 V_2''(y)+\frac{1}{2} C_9 V_2''(y)+\frac{3}{4} A_{122} V_2(y) V_2''(y)-\frac{1}{8} \hbar^2 A_{122}
	V_2{}^{(4)}(y) , \label{2 nonlinearODE for V_2}
	\end{align}                                                   
	where  $D_i $ and $\hat{D}_i$ are arbitrary integration constants.\\
Using \eqref{1 nonlinearODE for V_1}-\eqref{2 nonlinearODE for V_2} and the expressions of $f_{j2}$ and $f_{j4}$ already found, equation \eqref{D'7} reads
	\begin{align}
&\left(\frac{-3}{2}C_9 V_1(x)-C_8 V_1'(x)-x C_9 V_1'(x)+\beta _1'(x)+\frac{3}{4} h^2
A_{122} V_1''(x)\right)
+\left(\frac{3}{2} C_2 V_2(y)-C_5 V_2'(y)\nonumber \right.\\&+\left.
C_2 y V_2'(y)+\beta
_2'(y)-\frac{3}{4} h^2 A_{122} V_2''(y)\right)+\frac{1}{2} x^2 y^2
\left(D_3+K_3\right)+\frac{1}{2} x^2 y \left(D_2+2 \hat{K}_3\right)\nonumber \\&+
\frac{1}{2} x y^2
\left(K_2+2 \hat{D}_3\right)+x y \left(\hat{D}_2+\hat{K}_2\right)+x \hat{D}_1+y
\hat{K}_1+\frac{x^2 D_1}{2}+\frac{y^2 K_1}{2}=0. \label{Inbetween}	
	\end{align}
	By differentiating this equation three times with respect to $x$, we can integrate the resulting equation to obtain an expression for $\beta_1(x)$ in terms of the function $W_1(x)$. A similar calculation gives $\beta_2(y)$ in terms of $W_2(y)$:  
	\begin{subequations}
		\begin{align}
		\beta _1(x)=&+\frac{1}{2} W_1(x) C_9+C_{11}+x C_{12}+x^2 C_{13}+x^3 C_{14}+C_8 W_1'(x)+x C_9 W_1'(x)-\frac{3}{4} \hbar^2
		A_{122} W_1''(x),\label{Beta1}\\ 
		\beta _2(y)=&-\frac{1}{2} W_2(y) C_2+C_{15}+y C_{16}+y^2 C_{17}+y^3 C_{18}-y C_2 W_2'(y)+C_5 W_2'(y)+\frac{3}{4} \hbar^2 A_{122}
		W_2''(y),	\label{Beta2}	
		\end{align}
		where $C_i$ are arbitrary constants of integration.\\
	\end{subequations}
	With $\beta_1(x)$ and $\beta_2(y)$ given by \eqref{Beta1} and \eqref{Beta2}, equation \eqref{Inbetween} is now a second order polynomial in $x$ and $y$, since the functions $V_1(x)$ and $V_2(y)$ have all been canceled out. Setting its coefficients to zero, we find
	\begin{eqnarray}
	D_3=&-K_3,D_2=-2 \hat{K}_3,D_1=-6 C_{14},K_2=-2 \hat{D}_3,\hat{D}_2=-\hat{K}_2,\hat{D}_1=-2 C_{13},K_1=-6
	C_{18}, \nonumber\\&
	\hat{K}_1=-2 C_{17},C_{16}=-C_{12}. & \label{cont1}
	\end{eqnarray}
 We have now solved (with \eqref{1 nonlinearODE for V_1}-\eqref{2 nonlinearODE for V_2} satisfied) the system \eqref{D'1}-\eqref{D'8}. Substituting the expressions of $f_{j2}$ and $f_{j4}$ into the last equation \eqref{D'9}, we obtain
	\begin{align}
0=&\frac{1}{8} \hbar^4 A_{050} W_1{}^{(6)}(x)-\frac{5}{4} h^2 A_{050} W_1'(x) W_1{}^{(4)}(x)-\frac{5}{2} \hbar^2
A_{050} W_1''(x) W_1{}^{(3)}(x)\nonumber \\ &
+\frac{15}{4} A_{050} W_1'(x){}^2 W_1''(x)+\frac{1}{8} \hbar^4 A_{005}
W_2{}^{(6)}(y)-\frac{5}{4} \hbar^2 A_{005} W_2'(y) W_2{}^{(4)}(y)\nonumber \\ &
-\frac{5}{2} \hbar^2 A_{005} W_2''(y)
W_2{}^{(3)}(y)+\frac{15}{4} A_{005} W_2'(y){}^2 W_2''(y)+W_1''(x) \left(-6 \hbar^2 y C_4+2 \hbar^2 C_7-2 y C_{12}\nonumber \right.\\&+\left.
2C_{15}+2 y^2 C_{17}+2 y^3 C_{18}-2 y C_2 W_2'(y)+2 C_5 W_2'(y)+y C_5 W_2''(y)-\frac{1}{2} y^2 C_2
W_2''(y)\right)\nonumber \\ &
+W_2''(y) \left(6 \hbar^2 x C_4+2 \hbar^2 C_3+2 x C_{12}+2 C_{11}+2 x^2 C_{13}+2 x^3 C_{14}+2 x C_9
W_1'(x)+2 C_8 W_1'(x)\nonumber \right.\\&+\left.
x C_8 W_1''(x)+\frac{1}{2} x^2 C_9 W_1''(x)\right)+W_2'(y) W_2''(y) \bigg(3 x^3 C_4+3
C_1+3 x C_2+3 x^2 C_3+3 A_{023} W_1'(x)\nonumber\\&+\left.
3 x A_{122} W_1'(x)+\frac{3}{2} x A_{023} W_1''(x)+\frac{3}{4}
x^2 A_{122} W_1''(x)\right)+W_1'(x) W_1''(x) \left(-3 y^3 C_4+3 C_{10}\nonumber \right.\\&-\left.
3 y C_9+3 y^2 C_7+3 A_{032}
W_2'(y)-3 y A_{122} W_2'(y)+\frac{3}{2} y A_{032} W_2''(y)-\frac{3}{4} y^2 A_{122} W_2''(y)\right)\nonumber \\ &
+W_1(x)
\left(C_9 W_2''(y)+\frac{3}{2} A_{122} W_2'(y) W_2''(y)-\frac{1}{4} \hbar^2 A_{122}
W_2{}^{(4)}(y)\right)+W_2(y) \left(-C_2 W_1''(x)\nonumber \right.\\&-\left.
\frac{3}{2} A_{122} W_1'(x) W_1''(x)+\frac{1}{4} \hbar^2
A_{122} W_1{}^{(4)}(x)\right)+W_1{}^{(4)}(x) \left(\frac{1}{2} \hbar^2 y^3 C_4-\frac{1}{2} \hbar^2 y^2
C_7+\frac{1}{2} \hbar^2 y C_9\nonumber \right.\\&-\left.
\frac{\hbar^2 C_{10}}{2}-\frac{1}{2} h^2 A_{032} W_2'(y)+\frac{1}{2} \hbar^2 y A_{122}
W_2'(y)-\frac{1}{4} \hbar^2 y A_{032} W_2''(y)+\frac{1}{8} \hbar^2 y^2 A_{122} W_2''(y)\right)\nonumber \\ &
+W_2{}^{(4)}(y) \left(-\frac{1}{2} \hbar^2 x^3 C_4-\frac{1}{2} \hbar^2 x^2 C_3-\frac{1}{2} \hbar^2 x C_2-\frac{\hbar^2 C_1}{2}-\frac{1}{2} \hbar^2
A_{023} W_1'(x)-\frac{1}{2} \hbar^2 x A_{122} W_1'(x)\nonumber \right.\\&-\left.
\frac{1}{4} \hbar^2 x A_{023} W_1''(x)-\frac{1}{8} \hbar^2 x^2
A_{122} W_1''(x)\right). \label{D9''}
	\end{align} 
 We can now write the integral in \eqref{integral5-IIII} in terms of the constants $C_i$, $A_{ijk}$ and the functions $W_1(x)$ and $W_2(y)$. Its general but by no means final form is \\
\resizebox{.95\linewidth}{!}{
	\begin{minipage}{\linewidth}
		\begin{align}
		X=&A_{122}
		\left(-\frac{1}{2} \left\{p_1^3 p_2^2,y\right\}+\frac{3}{4} \left\{p_1^2 p_2,x W_2'(y)\right\}-\frac{1}{2}
		W_2(y) p_1^3-y W_2'(y) p_1^3\nonumber \right.\\&-\left.
		\frac{3}{8} \left\{p_1,W_2(y) W_1'(x)\right\}-\frac{3}{4} \left\{p_1,y W_1'(x)
		W_2'(y)\right\}+\frac{3}{4} h^2 W_2''(y) p_1+\frac{3}{16} \left\{p_1,x^2 W_2'(y) W_2''(y)\right\}\nonumber \right.\\&-\left.
		\frac{1}{32}
		\hbar^2 \left\{p_1,x^2 W_2{}^{(4)}(y)\right\}+\frac{1}{2} \left\{p_1^2 p_2^3,x\right\}-\frac{3}{4} \left\{p_1
		p_2^2,y W_1'(x)\right\}+\frac{1}{2} W_1(x) p_2^3+x W_1'(x) p_2^3\nonumber \right.\\&+\left.
		\frac{3}{8} \left\{p_2,W_1(x)
		W_2'(y)\right\}+\frac{3}{4} \left\{p_2,x W_1'(x) W_2'(y)\right\}-\frac{3}{4} \hbar^2 W_1''(x) p_2-\frac{3}{16}
		\left\{p_2,y^2 W_1'(x) W_1''(x)\right\}\nonumber \right.\\&+\left.
		\frac{1}{32} h^2 \left\{p_2,y^2 W_1{}^{(4)}(x)\right\}\right)\nonumber \\ &
		+A_{032} \left(p_1^3 p_2^2+\frac{3}{4} \left\{p_1 p_2^2,W_1'(x)\right\}+W_2'(y) p_1^3+\frac{3}{4} \left\{p_1,W_1'(x)
		W_2'(y)\right\}\nonumber \right.\\&+\left.
		\frac{3}{8} \left\{p_2,y W_1'(x) W_1''(x)\right\}-\frac{1}{16} \hbar^2 \left\{p_2,y
		W_1{}^{(4)}(x)\right\}\right)\nonumber \\ &
		+A_{023} \left(p_1^2 p_2^3+\frac{3}{4} \left\{p_1^2 p_2,W_2'(y)\right\}+W_1'(x)
		p_2^3+\frac{3}{4} \left\{p_2,W_1'(x) W_2'(y)\right\}\nonumber \right.\\&+\left.
		\frac{3}{8} \left\{p_1,x W_2'(y)
		W_2''(y)\right\}-\frac{1}{16} \hbar^2 \left\{p_1,x W_2{}^{(4)}(y)\right\}\right)\nonumber \\ &
		+A_{050} \left(p_1^5+\frac{5}{4}
		\left\{p_1^3,W_1'(x)\right\}+\frac{15}{16} \left\{p_1,W_1'(x){}^2\right\}+\frac{5}{16} \hbar^2
		\left\{p_1,W_1{}^{(3)}(x)\right\}\right)\nonumber \\ &
		+A_{005} \left(p_2^5+\frac{5}{4}
		\left\{p_2^3,W_2'(y)\right\}+\frac{15}{16} \left\{p_2,W_2'(y){}^2\right\}+\frac{5}{16} \hbar^2
		\left\{p_2,W_2{}^{(3)}(y)\right\}\right)\nonumber \\ &
		+C_1 \left(p_2^3+\frac{3}{4} \left\{p_2,W_2'(y)\right\}\right)+C_2
		\left(x p_2^3-\frac{1}{2} \left\{p_1 p_2^2,y\right\}-\frac{1}{2} W_2(y) p_1-y W_2'(y) p_1\nonumber \right.\\&+\left.
		\frac{3}{4} \left\{p_2,x W_2'(y)\right\}-\frac{1}{8} \left\{p_2,y^2 W_1''(x)\right\}\right)+C_3 \left(\frac{1}{2}
		\left\{p_1^2 p_2,y^2\right\}-\left\{p_1 p_2^2,x y\right\}+x^2 p_2^3\nonumber \right.\\&+\left.
		\frac{3}{4} \left\{p_2,x^2
		W_2'(y)\right\}-\frac{1}{4} \left\{p_2,x y^2 W_1''(x)\right\}+\frac{1}{4} \left\{p_1,x y^2
		W_2''(y)\right\}\right)+C_4 \left(-y^3 p_1^3\nonumber \right.\\&+\left.
		\frac{3}{2} \left\{p_1^2 p_2,x y^2\right\}-\frac{3}{2} \left\{p_1
		p_2^2,x^2 y\right\}+\frac{1}{2} x^3 p_2^3-\frac{3}{4} \left\{p_1,y^3 W_1'(x)\right\}+\frac{3}{4} \left\{p_2,x^3
		W_2'(y)\right\}\nonumber \right.\\&-\left.
		\frac{3}{8} \left\{p_2,x^2 y^2 W_1''(x)\right\}+\frac{3}{8} \left\{p_1,x^2 y^2
		W_2''(y)\right\}\right)+C_5 \left(p_1 p_2^2+p_1 W_2'(y)+\frac{1}{4} \left\{p_2,y W_1''(x)\right\}\right)\nonumber \\ &
		+C_6 \left(-\frac{1}{2} \left\{p_1^2 p_2,y\right\}+\frac{1}{2} \left\{p_1 p_2^2,x\right\}+\frac{1}{4} \left\{p_2,x y
		W_1''(x)\right\}-\frac{1}{4} \left\{p_1,x y W_2''(y)\right\}\right)\nonumber \\ &
		+C_7 \left(\frac{1}{2} \left\{p_1 p_2^2,x^2\right\}-\left\{p_1^2 p_2,x y\right\}+y^2 p_1^3+\frac{3}{4} \left\{p_1,y^2 W_1'(x)\right\}+\frac{1}{4}
		\left\{p_2,x^2 y W_1''(x)\right\}\nonumber \right.\\&-\left.
		\frac{1}{4} \left\{p_1,x^2 y W_2''(y)\right\}\right)+C_8 \left(p_1^2 p_2+p_2
		W_1'(x)+\frac{1}{4} \left\{p_1,x W_2''(y)\right\}\right)+C_9 \left(-y p_1^3 +\frac{1}{2} \left\{p_1^2
		p_2,x\right\}\nonumber \right.\\&+\left.
		\frac{1}{2} W_1(x) p_2+x W_1'(x) p_2-\frac{3}{4} \left\{p_1,y W_1'(x)\right\}+\frac{1}{8}
		\left\{p_1,x^2 W_2''(y)\right\}\right)
		+C_{10} \left(p_1^3+\frac{3}{4} \left\{p_1,W_1'(x)\right\}\right)\nonumber \\ &
		+C_{11}
		p_2+C_{12} \left(-y p_1+x p_2\right)+x^2 C_{13} p_2+x^3 C_{14} p_2 +C_{15} p_1+y^2 C_{17} p_1+y^3 C_{18} p_1. \label{Integ-Expli}
		\end{align}	
	\end{minipage}
}\\ \\
Further constraints on the coefficients $A_{ijk}$ and $C_i$ will be obtained below. We still have to solve eq. \eqref{1 nonlinearODE for V_1},...,\eqref{2 nonlinearODE for V_2} for $V_1(x)$ and $V_2(y)$ and assure compatibility with  
\eqref{D9''}. This will be done in Section $5$. The procedure will depend very much on the constants $A_{ijk}$ and $C_a$ in the integral \eqref{Integ-Expli}. Notice that each $A_{ijk}$ that remains free (i.e is not contained in the potential) provides an integral of order $5$. The constants $C_1,...C_{10}$ provide third order integrals, $C_{11},...C_{18}$ first order ones. We shall see that none of the order $1$ integrals survive. Some third order ones do, but are already known from earlier work  \cite{gravel2002superintegrability,gravel2004hamiltonians,marquette2017fourth}. 
\section{Calculation of the doubly exotic potentials}
	
	\subsection{$A_{122} \neq 0,A_{032}=0,A_{023}=0.$}
	If $A_{122} \neq0$ in the integral  \eqref{integral5-IIII}, we can assume that $A_{023}=A_{032}=0$ since we can get rid of these terms by two appropriate translations along $x$ and $y$ without affecting the separability of $V(x,y)$. In this case, the two ODEs \eqref{1 nonlinearODE for V_1} and \eqref{1 nonlinearODE for V_2} are no longer nonlinear and must be satisfied identically. Setting their coefficients to zero and using \eqref{cont1}, we find
	\begin{eqnarray}
	&C_6=0,\nonumber &\\
	C_7&=C_5=C_{17}=\hat{K}_1=\hat{K}_2=\hat{K}_3=0,&\nonumber \\
C_3&=C_8=C_{13}=\hat{D}_1=\hat{D}_2=\hat{D}_3=0,&\nonumber \\
&D_2=K_2=C_{17}=C_{13}=0.& \label{cont2}
	\end{eqnarray}
	In view of \eqref{cont1} and \eqref{cont2}, the remaining two nonlinear ODEs \eqref{2 nonlinearODE for V_1} ans \eqref{2 nonlinearODE for V_2} for $V_1(x)$ and $V_2(y)$ reduce to
	\begin{align}
	6 C_{18}-x^2 K_3-9 C_4 V_1(x)-9 x C_4 V_1'(x)-\frac{3}{4} A_{122} V_1'(x){}^2-\frac{1}{2} C_2 V_1''(x)-\frac{3}{2}
	x^2 C_4 V_1''(x)\nonumber \\
	-\frac{3}{4} A_{122} V_1(x) V_1''(x)
	+\frac{1}{8} \hbar^2 A_{122} V_1{}^{(4)}(x)=0, \label{Non-linear-ODE1}\\
	6 C_{14}+y^2 K_3+9 C_4 V_2(y)+9 y C_4 V_2'(y)+\frac{3}{4} A_{122} V_2'(y){}^2+\frac{3}{2} y^2 C_4
	V_2''(y)+\frac{1}{2} C_9 V_2''(y)\nonumber \\
	+\frac{3}{4} A_{122} V_2(y) V_2''(y)
	-\frac{1}{8} \hbar^2 A_{122} V_2{}^{(4)}(y)=0. \label{Non-linear-ODE2} 
	\end{align}
	At this point, equations \eqref{Non-linear-ODE1} and \eqref{Non-linear-ODE2} do not have the Painlev\'e property, for they do not pass the Painlev\'e test\cite{ablowitz1978nonlinear}. 
	Using the functions $W_1(x)$ and $W_2(y)$, we see that equations \eqref{Non-linear-ODE1} and \eqref{Non-linear-ODE2} admit two first integrals, namely\\
\begin{align}
6 x C_{18}-\frac{x^3 K_3}{3}-3 C_4 W_1(x)-6 x C_4 W_1'(x)-\frac{1}{2} C_2 W_1''(x)-\frac{3}{2} x^2
C_4 W_1''(x)\nonumber \\
-\frac{3}{4} A_{122} W_1'(x) W_1''(x)+\frac{1}{8} \hbar^2 A_{122} W_1{}^{(4)}(x)-C_{19}=0,\label{E1-integrated}\\
6 y C_{14}+\frac{y^3 K_3}{3}+3 C_4 W_2(y)+6 y C_4 W_2'(y)+\frac{3}{2} y^2 C_4 W_2''(y)+\frac{1}{2}
C_9 W_2''(y)\nonumber\\
+\frac{3}{4} A_{122} W_2'(y) W_2''(y)-\frac{1}{8} \hbar^2 A_{122} W_2{}^{(4)}(y)-C_{20}=0,\label{W1-integrated}
\end{align}
	where $C_{19}$ and $C_{20}$ are two integration constants. Next, the following combination
	\begin{align}
	\frac{\partial }{\partial y}\frac{\partial F}{\partial x}
	-\bigg( x^2 V_1''(x)+6 x V_1'(x)+6 V_1(x)\bigg)E2 -\bigg(y^2 V_2''(y)+6 y V_2'(y)+6 V_2(y)\bigg)E1\nonumber\\
	-\bigg(\frac{4 C_9 }{A_{122}}+\frac{12 y^2 C_4}{A_{122}}\bigg) E1
	-\bigg(\frac{12 x^2 C_4 }{A_{122}}+ \frac{4 C_2,}{A_{122}}\bigg)E2=0, \label{combi1}	
	\end{align}
	where $F$, $E1$ and $E2$ correspond to \eqref{D9''}, \eqref{Non-linear-ODE1} and \eqref{Non-linear-ODE2} respectively, gives a linear equation in $V_1(x)$ and $V_2(y)$, namely
	\begin{align}
&-\frac{24 C_2 C_{14}}{A_{122}}-\frac{72 x^2 C_4 C_{14}}{A_{122}}-\frac{72 y^2 C_4 C_{18}}{A_{122}}-\frac{24 C_9
		C_{18}}{A_{122}}-\frac{4 y^2 C_2 K_3}{A_{122}}+\frac{4 x^2 C_9 K_3}{A_{122}}
	+\biggl(-36 C_{14} \nonumber\\
	&-6 y^2K_3+\frac{108 y^2 C_4^2}{A_{122}}+\frac{36 C_4 C_9}{A_{122}}\biggr) V_1(x)+\left(-36 C_{18}+6 x^2 K_3-\frac{36 C_2 C_4}{A_{122}}-\frac{108 x^2 C_4^2}{A_{122}}\right) V_2(y)\nonumber\\
	&+\left(-36 x C_{14}
	-6 x y^2 K_3+\frac{108 x y^2 C_4^2}{A_{122}}+\frac{36 x C_4 C_9}{A_{122}}\right) V_1'(x)+\left(-36 y C_{18}+6 x^2 y K_3-\frac{36 y C_2C_4}{A_{122}}
	\right. \nonumber\\ 
	&-\left.\frac{108 x^2 y C_4^2}{A_{122}}\right) V_2'(y)
	+\left(-6 \hbar^2 C_4-2 C_{12}-6 x^2 C_{14}+6 y^2
	C_{18}-x^2 y^2 K_3+\frac{6 y^2 C_2 C_4}{A_{122}}\right. \nonumber\\ 
	&+\left.\frac{18 x^2 y^2 C_4^2}{A_{122}}+\frac{2 C_2 C_9}{A_{122}}
	+\frac{6 x^2 C_4 C_9}{A_{122}}\right) V_1''(x)
	+\biggl(6 \hbar^2 C_4+2 C_{12}+6 x^2 C_{14}-6 y^2
	C_{18}+x^2 y^2 K_3 \nonumber\\
	&-\frac{6 y^2 C_2 C_4}{A_{122}}-\frac{18 x^2 y^2 C_4^2}{A_{122}}
	-\frac{2 C_2 C_9}{A_{122}}
	-\frac{6 x^2 C_4 C_9}{A_{122}}\biggr) V_2''(y)=0. \label{combi1_expli}
	\end{align}
We use the derivatives of this equation to obtain linear ODEs for $V_1(x)$ and $V_2(y)$. Setting their coefficients to zero, we obtain further contraints on the constants, namely 
	\begin{equation}
	K_3=\frac{18 C_4^2}{A_{122}}, C_{18}=-\frac{C_2 C_4}{A_{122}}, C_{14}=\frac{C_4 C_9}{A_{122}}, C_{12}=\frac{C_2
		C_9-3 \hbar^2 C_4 A_{122}}{A_{122}}. \label{cont3}
	\end{equation}
	With the relations \eqref{cont3} verified, we find that the two ODEs \eqref{Non-linear-ODE1} and \eqref{Non-linear-ODE2} \textit{do} pass the Painlev\'e test. Indeed, we will see later that these two equations have the  Painlev\'e property.\\
	Next, we take the following combination:
	\begin{equation}
	F -x^2 W_1''(x) E4-y^2
	W_2''(y) E3-4 x W_1'(x) E4-4 y W_2'(y) E3=0, \label{combi2}
	\end{equation}
	where $F$, $E3$ and $E4$ correspond to \eqref{D9''}, \eqref{E1-integrated} and \eqref{W1-integrated} respectively. After a straightforward calculation, we obtain (after simplification using again \eqref{E1-integrated} and \eqref{W1-integrated} )
	\begin{align}
	 &\frac{4 x C_2 C_{20}}{A_{122}}+\frac{4 x^3 C_4 C_{20}}{A_{122}}+4 x C_{20} W_1'(x)+2 C_{20} W_1(x)+2 C_{15}
	W_1''(x)+x^2 C_{20} W_1''(x)\nonumber\\
	&+3 C_{10} W_1'(x) W_1''(x)+\frac{15}{4} A_{050} W_1(x)'{}^2 W_1''(x)-\frac{5}{2} \hbar^2
	A_{050} W_1'(x) W_1{}^{(3)}(x)-\frac{1}{2} \hbar^2 C_{10} W_1{}^{(4)}(x)\nonumber\\
	&-\frac{5}{4} \hbar^2 A_{050} W_1'(x)W_1{}^{(4)}(x)+\frac{1}{8} \hbar^4 A_{050} W_1{}^{(6)}(x) +\frac{4 y^3 C_4 C_{19}}{A_{122}}+\frac{4 y C_9 C_{19}}{A_{122}}+4 y C_{19} W_2'(y)\nonumber\\
	&+2 C_{19} W_2(y)+2 C_{11} W_2''(y)+y^2 C_{19} W_2''(y)+3 C_1 W_2'(y) W_2''(y) +\frac{15}{4} A_{005} W_2'(y){}^2 W_2''(y)\nonumber\\
	&-\frac{5}{2} \hbar^2 A_{005} W_2''(y) W_2'''(y)-\frac{1}{2} \hbar^2 C_1W_2{}^{(4)}(y)-\frac{5}{4} \hbar^2 A_{005} W_2'(y) W_2{}^{(4)}(y)+\frac{1}{8} \hbar^4 A_{005} W_2{}^{(6)}(y)=0. \label{combi_final}
	\end{align}
Terms depending on $x$ and those depending on $y$ can be separated. We set the $x$ dependent terms equal to a constant $-\kappa$ and the $y$ dependent ones equal to $+\kappa$. Thus the function $W_1(x)$ must satisfy the equation
\begin{align}
&\frac{4 C_4 C_{20} x^3}{A_{122}}+\frac{4 C_2 C_{20} x}{A_{122}}+\frac{1}{8} \hbar^4 A_{050}
W_1{}^{(6)}(x)-\frac{5}{4} \hbar^2 A_{050} W_1{}^{(4)}(x) W_1'(x)\nonumber \\&-\frac{5}{2} \hbar^2 A_{050} W_1{}^{(3)}(x)
W_1''(x)+\frac{15}{4} A_{050} W_1'(x){}^2 W_1''(x)-\frac{1}{2} C_{10} \hbar^2 W_1{}^{(4)}(x)+C_{20} x^2
W_1''(x)\nonumber \\&+3 C_{10} W_1'(x) W_1''(x)+4 C_{20} x W_1'(x)+2 C_{15} W_1''(x)+2 C_{20} W_1(x)=-\kappa, \label{painlevé5_V1}
	\end{align}
	where $\kappa$ is a constant. It must also satisfy \eqref{E1-integrated} with \eqref{cont3} taken into account:
\begin{align}
&-\frac{6 C_4^2 x^3}{A_{122}}-\frac{6 C_2 C_4 x}{A_{122}}+\frac{1}{8} \hbar^2 A_{122}
W_1{}^{(4)}(x)-\frac{3}{4} A_{122} W_1'(x) W_1''(x)-\frac{3}{2} C_4 x^2 W_1''(x)-6 C_4 x
W_1'(x)\nonumber \\&
-\frac{1}{2} C_2 W_1''(x)-3 C_4 W_1(x)-C_{19}=0. \label{Non-linear Painlevé1}
	\end{align}
A similar system must hold for $V_2(y)$, namely	
\begin{align}
&\frac{4 C_4 C_{19} y^3}{A_{122}}+\frac{4 C_9 C_{19} y}{A_{122}}+\frac{1}{8} \hbar^4 A_{005}
	W_2{}^{(6)}(y)-\frac{5}{4} \hbar^2 A_{005} W_2{}^{(4)}(y) W_2'(y)\nonumber \\&-\frac{5}{2} \hbar^2 A_{005} W_2{}^{(3)}(y)
	W_2''(y)+\frac{15}{4} A_{005} W_2'(y){}^2 W_2''(y)-\frac{1}{2} C_1 \hbar^2 W_2{}^{(4)}(y)+C_{19} y^2
W_2''(y)\nonumber\\&+3 C_1 W_2'(y) W_2''(y)+4 C_{19} y W_2'(y)+2 C_{11} W_2''(y)+2 C_{19} W_2(y)=\kappa, \label{painlevé5_V2}\\
\nonumber\\&
\frac{6 C_4^2 y^3}{A_{122}}+\frac{6 C_4 C_9 y}{A_{122}}-\frac{1}{8} \hbar^2 A_{122}
W_2{}^{(4)}(y)+\frac{3}{4} A_{122} W_2'(y) W_2''(y)+\frac{3}{2} C_4 y^2 W_2''(y)+6 C_4 y
W_2'(y)\nonumber \\&
+\frac{1}{2} C_9 W_2''(y)+3 C_4 W_2(y)-C_{20}=0 \label{Non-linear Painlevé2}.
	\end{align}
 We will solve these ODEs by distinguishing two sub-cases. We restrict ourselves to the ODEs satisfied by $V_1(x)$, for each solution satisfying \eqref{painlevé5_V1}  and \eqref{Non-linear Painlevé1} can be converted to a solution of the equations \eqref{painlevé5_V2}  and \eqref{Non-linear Painlevé2} by the following correspondence between the constants $C_i$ and $A_{i,j,k}$:

		\begin{equation*}
	\bigg(V_1(x), A_{122}, A_{050}, C_{10}, C_2, C_4, C_{15}, C_{19}, C_{20}, \kappa  \bigg)\\
		\end{equation*} 
		\begin{equation}
		\downarrow \label{Trans1}
		\end{equation}
				\begin{equation*}
	\bigg(V_2(y), -A_{122}, A_{005}, C_{1}, -C_9, -C_4, C_{11}, C_{20}, C_{19}, -\kappa \bigg)
	\end{equation*} 
	\subsubsection{Case A-1: $A_{050}=0$}
	In this case, the two ODEs that we need to solve are
	\begin{align}
-\frac{6 C_4^2 x^3}{A_{122}}-\frac{6 C_2 C_4 x}{A_{122}}+\frac{1}{8} h^2 A_{122}
W_1{}^{(4)}(x)-\frac{3}{4} A_{122} W_1'(x) W_1''(x)-\frac{3}{2} C_4 x^2 W_1''(x)-6 C_4 x
W_1'(x)\nonumber \\
-\frac{1}{2} C_2 W_1''(x)-3 C_4 W_1(x)-C_{19}=0, \label{A005: ODE1}\\  	
\frac{4 C_4 C_{20} x^3}{A_{122}}+\frac{4 C_2 C_{20} x}{A_{122}}-\frac{1}{2} C_{10} \hbar^2
W_1{}^{(4)}(x)+C_{20} x^2 W_1''(x)+3 C_{10} W_1'(x) W_1''(x)+4 C_{20} x W_1'(x)\nonumber \\ +2 C_{15} W_1''(x)+2
C_{20} W_1(x)+\kappa=0. \label{A005: ODE2}
	\end{align}
	Taking a linear combination of \eqref{A005: ODE1} and \eqref{A005: ODE2}, we obtain a linear ODE for $W_1(x)$, namely
	\begin{align}
&\kappa -\frac{24 x C_2 C_4 C_{10}}{A_{122}^2}-\frac{24 x^3 C_4^2 C_{10}}{A_{122}^2}-\frac{4 C_{10}
	C_{19}}{A_{122}}+\frac{4 x C_2 C_{20}}{A_{122}}+\frac{4 x^3 C_4 C_{20}}{A_{122}}\nonumber \\ &
+\left(2 C_{20}-\frac{12 C_4 C_{10}}{A_{122}}\right) W_1(x)+\left(4 x C_{20}-\frac{24 x C_4
	C_{10}}{A_{122}}\right) W_1'(x)\nonumber \\ &
+\left(2 C_{15}+x^2 C_{20}-\frac{2 C_2 C_{10}}{A_{122}}-\frac{6 x^2
	C_4 C_{10}}{A_{122}}\right) W_1''(x)=0.\label{nlODE}
	\end{align} 
	As usual the linear ODE \eqref{nlODE} must be satisfied identically and we must impose the following constraints
	\begin{equation}
	C_{20}=\frac{6 C_4 C_{10}}{A_{122}}, \ \ C_{15}=\frac{C_2 C_{10}}{A_{122}},\ \ \kappa =\frac{4 C_{10} C_{19}}{A_{122}}.
	\end{equation}
	The two nonlinear equations are now compatible, so we only need to solve \eqref{A005: ODE1}. We again distinguish $2$ subcases. 
\paragraph{$C_4=0$}	

	If $C_4=0$, eq.\eqref{A005: ODE1} can be integrated once:	
	\begin{equation}
	\frac{1}{8} \hbar^2 A_{122} W_1{}^{(3)}(x)-\frac{3}{8} A_{122} W_1'(x){}^2-\frac{1}{2} C_2 W_1'(x)-C_{19}
	x+K=0, \label{double-case}
	\end{equation}
where $K$ is an integration constant. Setting $W'_1(x)=V_1(x)=2 h^2 U_1(x)-\frac{2 C_2}{3 A_{122}}$ , this equation is transformed to 
	\begin{align}
	U_1''(x)=6 U_1(x){}^2+\frac{4 C_{19}}{\hbar^4 A_{122}}x-\left(\frac{2 C_2^2}{3 h^4
		A_{122}^2}+\frac{4 K}{\hbar^4 A_{122}}\right). \label{P1}
	\end{align} 
	The solution of \eqref{P1} is given by
	\begin{equation}
	U_1(x)=P_1(x,B_1,B_2),
	\end{equation}
where $B_1=\frac{4 C_{19}}{\hbar^4 A_{122}}$, $B_2=-\left(\frac{2 C_2^2}{3 \hbar^4
	A_{122}^2}+\frac{4 K}{\hbar^4 A_{122}}\right)$ and $P_1(x,B_1,B_2)$ satisfies the Painlev\'e-I equation 		
	\begin{equation}
	P_1''(x,B_1,B_2)=6 P_1(x,B_1,B_2){}^2+B_1 x+B_2. \label{PainlevéI-Def}
	\end{equation}
The potential reads
\begin{equation}
V_1(x)=2 \hbar^2 P_1(x,B_1,B_2)-\frac{2 C_2}{3 A_{122}}. \label{sol0}
\end{equation} 
Note that if $B_1=0$, the solution of \eqref{PainlevéI-Def} corresponds to the elliptic function of Weierstrass $\wp(x-x_0,g_1,g_2)$, with $g_1=-2B_1$ and $x_0, g_2$ are two arbitrary constants of integration. Otherwise, it is given by the first Painlev\'e transcendent function ($B_1$ can be scaled to $B_1=1$ and we can set $B_2=0$ by a translation).\\
\paragraph{$C_4\neq0$}	
When $C_4\neq0$, we differentiate \eqref{A005: ODE1} once and set
	\begin{equation}
	V_1(x)=2 \hbar^2 U_1(x)-\frac{2 C_4 x^2}{A_{122}}-\frac{2 C_2}{3 A_{122}}. \label{transfo1}
	\end{equation}
Then $U_1(x)$ must satisfy the following fourth order nonlinear equation of the polynomial type
	\begin{equation}
	U_1{}^{(4)}(x)=12 U_1'(x){}^2+12 U_1(x) U_1''(x)+\alpha  U_1'(x)+2 \alpha  U_1(x)-\frac{\alpha ^2 x^2}{6}, \label{F-I_1} 
	\end{equation}
	where $\alpha =\frac{24C_4}{\hbar^2 A_{122}}\neq0$.
	This ODE is well known, it has the Painlev\'e property and was derived from the point of view of Painlev\'e classification by Cosgrove in Ref.~\onlinecite{cosgrove2000higher} ( Eqs. (2.87) with $\beta=0$). It can also be obtained by a nonclassical reduction of the Boussinesq equation\cite {levi1989non} and the Kadomtsev-Petviashvili equation \cite{clarkson1991nonclassical}.\\
	We multiply \eqref{F-I_1} by the factor $x$ and  integrate once to give a member of Chazy Class XIII \cite{cosgrove2000chazy}. It can be integrated again to give the second order second degree equation $(19.7)$ in Ref.~\onlinecite{bureau1971equations}. Its solution \cite{cosgrove2000chazy} may be written in terms of the fourth Painlev\'e transcendent function, namely
	\begin{equation}
	U_1(x)=\frac{1}{2} \alpha_1P_4'(x,\alpha)-\frac{1}{2} \alpha P_4(x,\alpha){}^2-\frac{1}{2} \alpha x P_4(x,\alpha)-\frac{1}{6} \left(\frac{1}{2}\alpha x^2+K_1-\alpha_1\right),
	\end{equation}
	where $\alpha _1= \pm\sqrt{-\alpha}$ and $P_4(x,\alpha)=P_4(x,\alpha,K_1,K_2)$  satisfies the Painlev\'e-IV equation 
	\begin{equation}
	P_4(x,\alpha)''=\frac{\left(P_4(x,\alpha)'\right){}^2}{2 P_4(x,\alpha)}-\frac{3}{2}\alpha P_4(x,\alpha)^3-2\alpha x P_4(x,\alpha)^2-(\frac{1}{2}\alpha x^2+K_1)P_4(x,\alpha)+\frac{K_2}{P_4(x,\alpha)},\label{PainlebéIV- 1}
	\end{equation}
	with $K_1$ and $K_2$ two integration constants. The potential $V_1(x)$ reads 
	\begin{equation}
	V_1(x)=\frac{h^2}{12} \left(-3 x^2 \alpha -4 K_1+4 \alpha _1-12 x \alpha  P_4(x)-12 \alpha 
	P_4(x){}^2+12 \alpha _1 P_4'(x)\right)-\frac{2 C_2}{3 A_{122}}.\label{Solution1}
	\end{equation} 
	\subsubsection{Case A-2: $A_{050}\neq0$}
In this case we have to set $C_4=0$ in order to solve the system \eqref{painlevé5_V1}-\eqref{Non-linear Painlevé1}. Indeed, if $C_4\neq0$, $V_1(x)$ is given by \eqref{Solution1}. Therefore, by a straightforward but long calculation using Mathematica, we can use the second order Painlev\'e equation  \eqref{PainlebéIV- 1} to convert the derivative of \eqref{painlevé5_V1} into a first-order ODE for the fourth Painlev\'e transcendent function of the form
\begin{equation}
	x^2C_4 P_4(x,\alpha)^2P_4'(x,\alpha)=F(P_4(x,\alpha),P_4'(x,\alpha)),
\end{equation}
where $F$ is rational in $P_4(x,\alpha)$ and $P_4'(x,\alpha)$. But this is impossible since $P_4(x,\alpha)$ does not satisfy any first order ODE (in other words \eqref{painlevé5_V1} and \eqref{Non-linear Painlevé1} are incompatible for $C_4\neq0$).\\
We set $C_4=0$ and use \eqref{double-case} and its derivatives to remove all the nonlinear terms from \eqref{painlevé5_V1}. The linear ODE obtained reads 
\begin{align}
&\frac{4 C_2 C_{20}}{A_{122}}+6 C_{20} V_1(x)+\left(6 x C_{20}-\frac{3}{4} B_1 \hbar^2 A_{050}\right) V_1'(x)\nonumber \\
	&+\left(2
	C_{15}+x^2 C_{20}-\frac{1}{4} B_2 \hbar^2 A_{050}-\frac{1}{4} B_1 \hbar^2 x A_{050}+\frac{2 C_2^2
		A_{050}}{A_{122}^2}-\frac{2 C_2 C_{10}}{A_{122}}\right) V_1''(x)=0. \label{Compatibilité2 Subcase i}
\end{align}
Therefore, we must impose the following relations
\begin{align}
	\kappa =C_{20}=C_{19}=0,\ \ \
C_{15}=-\frac{C_2^2 A_{050}}{A_{122}^2}+\frac{C_2 C_{10}}{A_{122}}-\frac{K A_{050}}{A_{122}},
\end{align}
and the solution in this case is again given by \eqref{sol0} with $B_1=0$.
\subsection{$A_{122}=0$.}
We shall now repeat a similar analysis as in the previous case. When $A_{122}=0$, equations \eqref{2 nonlinearODE for V_1} and  \eqref{2 nonlinearODE for V_2} become linear in $V_1(x)$ and $V_2(y)$, so we must set
\begin{eqnarray}
&C_4=0,\nonumber &\\ 
&K_1=K_2=K_3=\hat{D_3}=C_3=C_2=0,\nonumber&\\
&D_1=D_2=D_3=\hat{K_3}=C_7=C_9=0.&
\end{eqnarray}
The remaining two nonlinear ODEs \eqref{1 nonlinearODE for V_1} and \eqref{1 nonlinearODE for V_2} read (after using \eqref{cont1})
\begin{align}
2 C_{17}-x \hat{K}_2+\frac{3}{2} C_6 V_1'(x)+\frac{3}{4} A_{032} V_1'(x){}^2+\frac{1}{2} C_5 V_1''(x)+\frac{1}{2} x
C_6 V_1''(x)+\frac{3}{4} A_{032} V_1(x) V_1''(x)\nonumber\\
-\frac{1}{8} \hbar^2 A_{032} V_1{}^{(4)}(x)=0, \label{A122=0,Non-linear-ODE1}\\ 	
2 C_{13}+y \hat{K}_2-\frac{3}{2} C_6 V_2'(y)+\frac{3}{4} A_{023} V_2'(y){}^2-\frac{1}{2} y C_6 V_2''(y)+\frac{1}{2}
C_8 V_2''(y)+\frac{3}{4} A_{023} V_2(y) V_2''(y)\nonumber\\
-\frac{1}{8} \hbar^2 A_{023} V_2{}^{(4)}(y)=0. \label{A122=0,Non-linear-ODE2} 
\end{align}                                          
Unlike the case where $A_{122}$ did not vanish, these two equations can be integrated without any use of the auxiliary functions $W_1(x)$ and $W_2(y)$:
\begin{align}
2 x C_{17}-\frac{1}{2} x^2 \hat{K}_2+C_6 V_1(x)+\frac{1}{2} C_5 V_1'(x)+\frac{1}{2} x C_6 V_1'(x)+\frac{3}{4}
A_{032} V_1(x) V_1'(x)\nonumber\\
-\frac{1}{8} \hbar^2 A_{032} V_1{}^{(3)}(x)-C_{21}=0, \label{A122=0,Non-linear-ODE1Integr}\\
2 y C_{13}+\frac{1}{2} y^2 \hat{K}_2-C_6 V_2(y)-\frac{1}{2} y C_6 V_2'(y)+\frac{1}{2} C_8 V_2'(y)+\frac{3}{4}
A_{023} V_2(y) V_2'(y)\nonumber\\
-\frac{1}{8} \hbar^2 A_{023} V_2{}^{(3)}(y)-C_{22}=0,\label{A122=0,Non-linear-ODE2Integr}
\end{align}
where $C_{21}$ and $C_{22}$ are integration constants. Next, we use the combination
\begin{equation}
\frac{\partial }{\partial y}\frac{\partial F}{\partial x}-2 x V_1''(x) \hat{E2}-6 V_1'(x)
\hat{E2}-2 y V_2''(y)\hat{E1}-6 V_2'(y) \hat{E1}=0,
\end{equation}
where $F$,$\hat{E1}$ and $\hat{E2}$ correspond to \eqref{D9''}, \eqref{A122=0,Non-linear-ODE1} and \eqref{A122=0,Non-linear-ODE2} respectively  to obtain the following linear equation
\begin{align}
&\left(-12 C_{13} V_1'(x)-4 x C_{13} V_1''(x)-2 C_{12} V_1''(x)\right)-12 C_{17} V_2'(y)-4 y C_{17} V_2''(y)+2 C_{12}
V_2''(y)\nonumber\\
&+x \left(6 \hat{K}_2 V_2'(y)+4 C_{13} V_2''(y)+2 y \hat{K}_2 V_2''(y)\right)-y \left(6 \hat{K}_2 V_1'(x)-4
C_{17} V_1''(x)+2 x \hat{K}_2 V_1''(x)\right)=0. \label{A122=0, combi1expli}
\end{align}
We use the derivatives of \eqref{A122=0, combi1expli} to obtain linear ODEs for $V_1(x)$ and $V_2(y)$. Setting their coefficients to zero, we obtain 
\begin{equation}
\hat{K}_2=C_{17}=C_{12}=C_{13}=0.
\end{equation}
We use again the combination
\begin{equation}
F-2 x V_1'(x) \hat{E3}-4 V_1(x) \hat{E3}-2 y V_2'(y) \hat{E4}-4 V_2(y) \hat{E4}=0,
\end{equation}
where $F$,$\hat{E1}$ and $\hat{E2}$ correspond to \eqref{D9''}, \eqref{A122=0,Non-linear-ODE1Integr} and \eqref{A122=0,Non-linear-ODE2Integr} to obtain the following separable equation (in this case we can express $F$ in terms of $V_1(x)$ and $V_2(y)$)
\begin{align}
&4 C_{22} V_1(x)+2 C_{15} V_1'(x)+2 x C_{22} V_1'(x)+3 C_{10} V_1(x) V_1'(x)+\frac{15}{4}
A_{050} V_1(x){}^2 V_1'(x)\nonumber\\
&-\frac{5}{2} \hbar^2 A_{050} V_1'(x) V_1''(x) -\frac{1}{2}
\hbar^2 C_{10} V_1{}^{(3)}(x)-\frac{5}{4} \hbar^2 A_{050} V_1(x) V_1{}^{(3)}(x)+\frac{1}{8}
\hbar^4 A_{050} V_1{}^{(5)}(x)\nonumber\\
&+4 C_{21} V_2(y)+2 C_{11} V_2'(y)+2 y C_{21} V_2'(y) +3C_1 V_2(y) V_2'(y)+\frac{15}{4} A_{005} V_2(y){}^2 V_2'(y)\nonumber\\
&-\frac{5}{2} \hbar^2
A_{005} V_2'(y) V_2''(y)-\frac{1}{2} \hbar^2 C_1 V_2{}^{(3)}(y)-\frac{5}{4} \hbar^2 A_{005} V_2(y) V_2{}^{(3)}(y) +\frac{1}{8} \hbar^4 A_{005} V_2{}^{(5)}(y)=0. \label{Deriv}
\end{align}
Thus, $V_1(x)$ must satisfy the system
\begin{align}
&4 C_{22} V_1(x)+2 C_{15} V_1'(x)+2 x C_{22} V_1'(x)+3 C_{10} V_1(x) V_1'(x)+\frac{15}{4}
A_{050} V_1(x){}^2 V_1'(x)\nonumber\\&
-\frac{5}{2} \hbar^2 A_{050} V_1'(x) V_1''(x) -\frac{1}{2}
\hbar^2 C_{10} V_1{}^{(3)}(x)-\frac{5}{4} \hbar^2 A_{050} V_1(x) V_1{}^{(3)}(x)+\frac{1}{8}
\hbar^4 A_{050} V_1{}^{(5)}(x)=\kappa, \label{PainlevéIV-2}\\
&-C_{21}+C_6 V_1(x)+\frac{1}{2} C_5 V_1'(x)+\frac{1}{2} x C_6 V_1'(x)+\frac{3}{4} A_{032} V_1(x)
V_1'(x)-\frac{1}{8} \hbar^2 A_{032} V_1{}^{(3)}(x)=0.\label{F-I_2}
\end{align}
Eq. \eqref{PainlevéIV-2} is obtained by taking the derivative $\frac{\partial}{\partial x}$ of \eqref{Deriv}, \eqref{F-I_2} is a consequence of \eqref{A122=0,Non-linear-ODE1Integr}.\\    
Similarly, $V_2(y)$ must satisfy 
\begin{align}
&4 C_{21} V_2(y)+2 C_{11} V_2'(y)+2 y C_{21} V_2'(y)+3 C_1 V_2(y) V_2'(y)+\frac{15}{4} A_{005}
V_2(y){}^2 V_2'(y)\nonumber \\ &-\frac{5}{2} \hbar^2 A_{005} V_2'(y) V_2''(y)-\frac{1}{2} \hbar^2 C_1
V_2{}^{(3)}(y)-\frac{5}{4} \hbar^2 A_{005} V_2(y) V_2{}^{(3)}(y)+\frac{1}{8} \hbar^4 A_{005} V_2{}^{(5)}(y)=-\kappa ,\\
&-C_{22}-C_6 V_2(y)+\frac{1}{2} C_8 V_2'(y)-\frac{1}{2} y C_6 V_2'(y)+\frac{3}{4} A_{023} V_2(y)
V_2'(y)-\frac{1}{8} \hbar^2 A_{023} V_2{}^{(3)}(y)=0.
\end{align}
We shall again restrict ourselves to the solutions of \eqref{PainlevéIV-2} and \eqref{F-I_2}. To obtain the corresponding solutions of $V_2(y)$, we set 	
\begin{equation*}
\bigg(V_1(x),A_{050},C_5,C_6,C_{10},C_{15},C_{21},C_{22},\kappa  \bigg)
\end{equation*} 
\begin{equation}
\downarrow
\end{equation} 
\begin{equation*}
\bigg(V_2(y),A_{005},C_8,-C_6,C_{1},C_{11},C_{22},C_{21},-\kappa  \bigg) \label{Trans2}
\end{equation*} 

\subsubsection{Case B-1: $A_{050}=0$, $A_{032}\neq0$}
Since $A_{032}\neq0$, we use equation \eqref{F-I_2} to eliminate the nonlinear terms from \eqref{PainlevéIV-2}. The constraints on the resulting linear equation guarantee the compatibility of the system and we only need to solve \eqref{F-I_2}. Indeed, we obtain 
\begin{equation}
C_{22}=\frac{C_6 C_{10}}{A_{032}}, C_{15}=\frac{C_5 C_{10}}{A_{032}},\kappa=\frac{4 C_{10} C_{21}}{A_{032}}.
\end{equation}
If we set
\begin{equation}
V_1(x)=2 \hbar^2 U_1(x)-\frac{2 C_6  x}{3 A_{032}}-\frac{2 C_5 }{3 A_{032}},
\end{equation}
the function $U_1(x)$ must satisfy
\begin{equation}
U_1{}^{(3)}(x)=12 U_1(x) U_1'(x)-\frac{\beta ^2 x}{6}+\beta  U_1(x)+\mu, \label{Chazy XIII}
\end{equation}
where $\beta=\frac{4 C_6}{\hbar^2 A_{032}}$ and $\mu =-\frac{8 C_5 C_6}{3 \hbar^4 A_{032}^2}-\frac{4 C_{21}}{\hbar^4 A_{032}}$.\\
Equation \eqref{Chazy XIII} appears in Ref.~\onlinecite{cosgrove2000chazy} (eq.Chazy-XIII.b) as a member of the Chazy equations in the third-order polynomial class. It has the Painlev\'e property and can be obtained by each of the three group-invariant reductions of the KdV equation. When $\beta\neq0$, its solution is given by
\begin{equation}
U_1(x)=\frac{1}{2} \left(\epsilon _1 P_2'(x,\beta)+P_2(x,\beta){}^2\right)+\frac{1}{12} \left(\beta x+\gamma\right), \label{finalcase}
\end{equation}
where $ \epsilon _1 = \pm 1 $ and $P_2(x,\beta)=P_2(x,\beta,\gamma,\delta)$ satisfies the Painlev\'e-II equation
\begin{equation}
P_2''(x,\beta)=2 P_2(x,\beta){}^3+\left(\beta x+\gamma\right) P_2(x,\beta)+\delta, \label{PainlevéReduce}
\end{equation}
where $\delta$ is a constant of integration and $\gamma:= -6\mu/\beta.$\\
In this case, $V_1(x)$ reads
\begin{equation}
V_1(x)= -\frac{2 C_5}{3 A_{032}}+\hbar^2 \left(P_2(x,\beta ){}^2+\epsilon _1 P_2'(x,\beta
)-\frac{\mu }{\beta }\right). \label{Solution Reduce}
\end{equation}
If $\beta=0$, $U_1(x)$ is given by
\begin{equation}
U_1(x)=P_1(x,\mu,K),
\end{equation}
$K$ being an integration constant.
 \subsubsection{Case B-2: $A_{050}\neq0$, $A_{032}=0$}
 In this case, the linearity of equation \eqref{F-I_2} implies $C_5=C_6=C_{21}=0$. Consequently, $V_1(x)$ must only satisfy \eqref{PainlevéIV-2} which, by setting $V_1(x)=2 \hbar^2 U_1(x)-\frac{2 C_{10}}{5 A_{050}}$, we transform to 
 \begin{multline}
 U_1{}^{(5)}(x)=20 U_1(x) U_1{}^{(3)}(x)+40 U_1'(x) U_1''(x)-120 U_1(x){}^2
 U_1'(x)+(\lambda  x+\alpha ) U_1'(x)+2 \lambda  U_1(x)+\omega, \label{FifIII}
 \end{multline}
 where 
 \begin{equation}
\lambda =-\frac{16 C_{22}}{\hbar^4 A_{050}},\ \ \alpha =\frac{24
 	C_{10}^2}{5 \hbar^4 A_{050}^2}-\frac{16 C_{15}}{\hbar^4 A_{050}},\ \
 \omega =\frac{32 C_{10}
 	C_{22}}{5 \hbar^6 A_{050}^2}+\frac{4 \kappa }{\hbar^6 A_{050}}.
 \end{equation}
 This equation passes the Painlev\'e test and appears in the list of fifth-order Painlev\'e type equations of polynomial class in Ref.~\onlinecite{cosgrove2000higher} as the equation Fif-III. We do not know the general solution of this ODE, but we shall see that it admits a special solution in terms of the first Painlev\'e transcendent.\\
 If $\lambda=0$, equation \eqref{FifIII} admits the first integral,
 \begin{equation}
 U_1{}^{(4)}(x)=20 U_1(x) U_1''(x)+10 U_1'(x){}^2-40 U_1(x){}^3+x \omega +\alpha  U_1(x)+\gamma. \label{F-V}
 \end{equation}
 This equation has the Painlev\'e property and appears also in the list of fourth-order Painlev\'e type equations of polynomial class in Ref.~\onlinecite{cosgrove2000higher}. If $\omega\neq0$, we do not know any exact solutions of this equation either. It is possible that, in this case,  it defines a new transcendent, that is it has no elementary solution expressible in terms of known transcendents (including the six Painlev\'e transcendents). The case $\omega=0$ can be solved in terms of hyperelliptic functions. We refer the reader to Ref.~\onlinecite{cosgrove2000higher} for the details. If we set $\alpha=\gamma=0$ in \eqref{F-V}, we find the special case obtained in Ref.~\onlinecite{gungor2017heisenberg}.\\
 When $\lambda\neq0$, equation \eqref{FifIII} admits the first integral
 \begin{equation}
 2HH''-(H')^2-(8U_1(x)+4\omega/\lambda)H^2+\tilde{K}=0,
 \end{equation}
 where $H$ is the auxiliary variable
 \begin{equation}
 H:= U_1''(x)-6U_1(x)^2+4(\omega/ \lambda)U_1(x)+\frac{1}{4}(\lambda x+\alpha)-4( \omega/ \lambda )^2.
 \end{equation}
 When $\tilde{K}=0$, a particular solution can be obtained by setting $H=0$. Therefore, this solution may be written in terms of the first Painlev\'e transcendent, namely
  \begin{equation}
  	U_1(x)=P_1(x,B_1,B_2)-\frac{1}{3} (\omega/\lambda),
  \end{equation}
  where $B_1=-\frac{\lambda}{4}$ and $B_2=-\frac{\alpha}{4}+\frac{10}{3}(\omega/\lambda)$.
  The potential $V_1(x)$ reads 
  \begin{equation}
  V_1(x)=2 \hbar^2 P_1(x,B_1,B_2)-\frac{2 \hbar^2}{3} (\omega/\lambda)-\frac{2 C_{10}}{5 A_{050}}. \label{waw2}
  \end{equation}
  \subsubsection{Case B-3: $A_{050}\neq0$, $A_{032}\neq0$}
In this case, if we assume that $C_6\neq0$, we can substitute $V_1(x)$ given in \eqref{Solution Reduce} into \eqref{PainlevéIV-2} and then reduce the order of the resulting equation using  \eqref{PainlevéReduce} and its derivatives. We obtain a first order ODE for $P_2(x)$ of the form
\begin{equation}
x^2C_6P_2(x,\beta)P_2'(x,\beta)=F(P_2'(x,\beta),P_2(x,\beta)),
\end{equation}
where $P_2(x)$ is the second Painlev\'e transcendent and $F$ is polynomial in $P_2(x)$ and $P_2'(x)$. Since this is impossible, we must set $C_6=0$. Consequently, we integrate \eqref{F-I_2} to 
\begin{equation}
C_{23}-x C_{21}+\frac{1}{2} C_5 V_1(x)+\frac{3}{8} A_{032} V_1(x){}^2-\frac{1}{8} h^2
A_{032} V_1''(x)=0. \label{final_integtwice}
\end{equation}
We use this equation to reduce the order of \eqref{PainlevéIV-2} and obtain the linear ODE,
\begin{align}
&-\kappa +\frac{4 C_{10} C_{21}}{A_{032}}-\frac{4 C_5 C_{21}
	A_{050}}{A_{032}^2}+\left(4 C_{22}+\frac{4 C_{21} A_{050}}{A_{032}}\right)
V_1(x)\nonumber \\&
+\left(2 C_{15}+2 x C_{22}-\frac{2 C_5 C_{10}}{A_{032}}+\frac{2 C_5^2 A_{050}}{A_{032}^2}+\frac{2 x C_{21} A_{050}}{A_{032}}-\frac{2 C_{23}
	A_{050}}{A_{032}}\right) V_1'(x)=0.
\end{align}
As usual, we must set
\begin{equation}
C_{22}=-\frac{C_{21}A_{050}}{A_{032}}, \ \ C_{15}=\frac{C_5 C_{10}}{A_{032}}-\frac{C_5^2
	A_{050}}{A_{032}^2}+\frac{C_{23} A_{050}}{A_{032}}, \ \
 \kappa =\frac{4 C_{10} C_{21} }{A_{032}}-\frac{4 C_5 C_{21} A_{050}}{A_{032}^2} \label{cont_waw}.
\end{equation}
$V_1(x)$ is then given by
\begin{equation}
V_1(x)=2 \hbar^2 P_1(x,B_1,B_2)-\frac{2 C_5}{3 A_{032}},
\end{equation}
where $B_1=-\frac{4C_{21}}{\hbar^4 A_{032}}$ and $B_2=-\frac{1}{\hbar^4 A_{032}}(\frac{2 C_5^2}{3
A_{032}}-4 C_{23}).$ 
 \subsubsection{Case B-4: $A_{050}=0$, $A_{032}=0$}
 From equation \eqref{F-I_2}, we must set
 \begin{equation}
 C_5=C_6=0,
 \end{equation}
 and the only ODE that we need to solve reads
 \begin{equation}
 4 C_{22} V_1(x)+2 C_{15} V_1'(x)+2 x C_{22} V_1'(x)+3 C_{10} V_1(x) V_1'(x)-\frac{1}{2}
 \hbar^2 C_{10} V_1{}^{(3)}(x)=\kappa. \label{CaseB4-eq}
 \end{equation}
If $C_{10}$=0, \eqref{CaseB4-eq} is linear and we must have $C_{15}=C_{22}=\kappa=0$. \\
For $C_{10}\neq0$, we set
\begin{equation}
V_1(x)=2 \hbar^2 U_1(x)-\frac{2 C_{22} x+2 C_{15}}{3 C_{10}},
\end{equation}
and $U_1(x)$ must satisfy
\begin{equation}
U_1{}^{(3)}(x)=12 U_1(x) U_1'(x)-\frac{\beta ^2 x}{6}+\beta  U_1(x)+\mu,
\end{equation}
where 
\begin{equation}
\beta =\frac{4 C_{22}}{\hbar^2 C_{10}}, \ \ \ \mu =-\frac{\kappa }{\hbar^4 C_{10}}-\frac{8 C_{15}
	C_{22}}{3 \hbar^4 C_{10}^2}.
\end{equation}
This case is similar to the case B-1 and the solution is given by \eqref{finalcase}.
\section{Summary of the results}
In Section $5$ the results are ordered by the form of the integral of motion \eqref{Integ-Expli} and we specified all the constants in the integral and in the potentials $V_1(x)$ and $V_2(y)$. We found all exotic potentials $V_1(x)$ and by symmetry (see e.g. \eqref{Trans1}) also $V_2(y)$. Certain potentials $V= V_1(x)+V_2(y)$ appeared more than once with different integrals of motion. In this section we will order the results according to the form of the potentials and for each of the $9$ classes of potentials list all integrals. In each class we have the integrals $\mathcal{H}_1$ and $\mathcal{H}_2$ of \eqref{Super-Integrals} and at least one integral of order $5$ and hence of the form $X$ of eq. \eqref{Integ-Expli}. The question of their algebraic independence was not yet discussed. In classical mechanics at most $3$ such integrals in $E_2$ can be functionally independent (in $E_n$ the number is $2n-1$). In quantum mechanics no such theorems are available. We can however make use of the results of Burchnall-Chaundy theory concerning commutative ordinary differential operators \cite{burchnall1923commutative,mironov2012commuting}. A relevant result is:
\begin{theorem}[Burchnall-Chaundy]
	\label{B-C}
Consider two operators
\begin{equation}
X_1= \partial _x^n+ \sum_{i=0}^{n-2} u_i(x)\partial_x^i, \ \ \
X_2= \partial _x^m+ \sum_{i=0}^{m-2} v_i(x)\partial_x^i. \label{BC}
\end{equation}
Then  $[X_1,X_2]=0$ implies that there exists a nonzero polynomial $P$ such that $P(X_1,X_2)=0$. In other words, $X_1$ and $X_2$ are not polynomially independent.
\end{theorem}
In our case the role of $X_1$ in \eqref{BC} is played by $\mathcal{H}_1=-\partial _x^2 + V_1(x)$ (or $\mathcal{H}_2$) and $X_2$ by one of the operators commuting with $\mathcal{H}$ found in section $5$.\\
Let us run through all potentials found above: they are candidates for being superintegrable.\\
$Q_1:$
\begin{empheq}[box=\fbox]{align}
V_1(x)= 2\hbar^2 \wp(x,g_1,g_2), \ \ \ V_2(y)=2\hbar^2\wp(y,\hat{g_1},\hat{g_2}).
\end{empheq}
\begin{align}
X_{122}=& \left.-
\frac{1}{2} \left\{p_1^3 p_2^2,y\right\}+\frac{3}{4} \left\{p_1^2 p_2,x W_2'(y)\right\}
-\frac{1}{2} W_2(y)
p_1^3-y W_2'(y) p_1^3-\frac{3}{8} \left\{p_1,W_2(y) W_1'(x)\right\}\nonumber \right.\\&-\left.
\frac{3}{4}
\left\{p_1,y W_1'(x) W_2'(y)\right\}+\frac{3}{4} \hbar^2 p_1 W_2''(y)+\frac{3}{16}
\left\{p_1,x^2 W_2'(y) W_2''(y)\right\}\nonumber \right.\\&-\left.
\frac{1}{32} \hbar^2 \left\{p_1,x^2
W_2{}^{(4)}(y)\right\}+\frac{1}{2} \left\{p_1^2 p_2^3,x\right\}-\frac{3}{4}
\left\{p_1 p_2^2,y W_1'(x)\right\}+\frac{1}{2} W_1(x) p_2^3+x W_1'(x)
p_2^3\nonumber \right.\\&+\left.
\frac{3}{8} \left\{p_2,W_1(x) W_2'(y)\right\}+\frac{3}{4} \left\{p_2,x W_1'(x)
W_2'(y)\right\}-\frac{3}{4} \hbar^2 W_1''(x) p_2\nonumber \right.\\&-\left.
\frac{3}{16} \left\{p_2,y^2 W_1'(x)
W_1''(x)\right\}
+\frac{1}{32} \hbar^2 \left\{p_2,y^2 W_1{}^{(4)}(x)\right\}\nonumber \right. ,\\
X_{032}=&\left. \left(p_2^2+W_2'(y)\right)\left( p_1^3+\frac{3}{4} \left\{p_1,W_1'(x)\right\} \right)\right.=\mathcal{H}_2 X_A ,\nonumber\\
X_{023}= &\left.\left(p_1^2+W_1'(x)\right) \left( p_2^3+\frac{3}{4} \left\{p_2,W_2'(y)\right\} \right)\right.=\mathcal{H}_1 X_B ,\nonumber\\
X_{050}=&p_1^5+\frac{5}{4}  \left\{p_1^3,W_1'(x)\right\}+\frac{15}{16}
\left\{p_1,W_1'(x){}^2\right\}+\frac{5}{16} \hbar^2
\left\{p_1,W_1{}^{(3)}(x)\right\}-\frac{g_1 \hbar^2}{16}, \label{Q_1} \\
X_{005}=&p_2^5+\frac{5}{4} \left\{p_2^3,W_2'(y)\right\}+\frac{15}{16}
\left\{p_2,W_2'(y){}^2\right\}+\frac{5}{16} \hbar^2
\left\{p_2,W_2{}^{(3)}(y)\right\}-\frac{1}{16} \hat{g}_1 \hbar^2 ,\nonumber\\
X_{A}=& p_1^3+\frac{3}{4} \left\{p_1,W_1'(x)\right\} ,\nonumber\\
X_{B}=&p_2^3+\frac{3}{4} \left\{p_2,W_2'(y)\right\}.\nonumber
\end{align}
We have $7$ linearly independent operators commuting with $\mathcal{H}$, given in \eqref{Q_1}. By Theorem \ref{B-C} we see that $X_A$ and $X_{050}$ are polynomially related with $\mathcal{H}_1$, $X_B$ and $X_{005}$  with $\mathcal{H}_2$. The integrals $X_{032}$ and $X_{023}$ are simply products of lower order integrals.\\
The 3 polynomially independent integrals are $\mathcal{H}_1$, $\mathcal{H}_2$ and $X_{122}$ so the system $Q_1$ involving two Weierstrass elliptic functions is superintegrable.
\begin{remark}
the notations used here and below are that e.g $X_{005}$ corresponds to setting $A_{005}=1$, all other $A_{ijk}=0$ except those that are proportional to $A_{005}$. Lower order integrals are listed separately (e.g. $X_A$).  	
\end{remark}
$Q_2:$
\begin{empheq}[box=\fbox]{align}
V_1(x)=&\hbar^2\bigg(\alpha_1 P_4'(x,\alpha)-x\alpha P_4(x,\alpha)-\alpha P_4(x,\alpha)^2-\frac{\alpha x^2}{4} \bigg),\nonumber \\
V_2(y)=&\hbar^2\bigg(\alpha_1 P_4'(y,\alpha)-y\alpha P_4(y,\alpha)-\alpha P_4(y,\alpha)^2-\frac{\alpha y^2}{4} \bigg).
\end{empheq}
where $\alpha_1=\pm\sqrt{\alpha}\neq0$ and $P_4(x,\alpha)=P_4(x,\alpha,K_1,K_2)$, $P_4(y,\alpha)=P_4(y,\alpha, \hat{K_1}, \hat{K_2})$ are the fourth Painlev\'e transcendents  given by \eqref{PainlebéIV- 1}.
\begin{flalign}
X_{122}=& \left.-
\frac{1}{2} \left\{p_1^3 p_2^2,y\right\}+\frac{3}{4} \left\{p_1^2 p_2,x W_2'(y)\right\}
-\frac{1}{2} W_2(y)
p_1^3-y W_2'(y) p_1^3-\frac{3}{8} \left\{p_1,W_2(y) W_1'(x)\right\}\nonumber \right.\\&-\left.
\frac{3}{4}
\left\{p_1,y W_1'(x) W_2'(y)\right\}+\frac{3}{4} \hbar^2 p_1 W_2''(y)+\frac{3}{16}
\left\{p_1,x^2 W_2'(y) W_2''(y)\right\}\nonumber \right.\\&-\left.
\frac{1}{32} \hbar^2 \left\{p_1,x^2
W_2{}^{(4)}(y)\right\}+\frac{1}{2} \left\{p_1^2 p_2^3,x\right\}-\frac{3}{4}
\left\{p_1 p_2^2,y W_1'(x)\right\}+\frac{1}{2} W_1(x) p_2^3+x W_1'(x)
p_2^3\nonumber \right.\\&+\left.
\frac{3}{8} \left\{p_2,W_1(x) W_2'(y)\right\}+\frac{3}{4} \left\{p_2,x W_1'(x)
W_2'(y)\right\}-\frac{3}{4} \hbar^2 W_1''(x) p_2\nonumber \right.\\&-\left.
\frac{3}{16} \left\{p_2,y^2 W_1'(x)
W_1''(x)\right\}
+\frac{1}{32} \hbar^2 \left\{p_2,y^2 W_1{}^{(4)}(x)\right\}\nonumber \right.\\&
+\left. \frac{1}{24} \text{$\alpha \hbar $}^2 \left(-y^3 p_1^3+\frac{3}{2}
\left\{p_1^2 p_2,x y^2\right\}-\frac{3}{4} \left\{p_1,y^3 W_1'(x)\right\}+\frac{3}{8}
\left\{p_1,x^2 y^2 W_2''(y)\right\}\nonumber \right. \right.\\&+\left.\left.
3 \hbar^2 y p_1+x^3 p_2^3-\frac{3}{2} \left\{p_1
p_2^2,x^2 y\right\}+\frac{3}{4} \left\{p_2,x^3 W_2'(y)\right\}-\frac{3}{8}
\left\{p_2,x^2 y^2 W_1''(x)\right\}-3 \hbar^2 x p_2\right) \nonumber \right.\\&+\left.
\frac{1}{2} \hbar^2 \left(\hat{K_1}-\alpha _1\right) \left(-y p_1^3+\frac{1}{2} \left\{p_1^2
p_2,x\right\}+\frac{1}{2} W_1(x) p_2 +
x W_1'(x) p_2-\frac{3}{4} \left\{p_1,y
W_1'(x)\right\}\nonumber \right.\right.&\\&+\left.\left.\frac{1}{8} \left\{p_1,x^2 W_2''(y)\right\}+\frac{1}{2} \hbar^2 y p_1
\left(K_1-\alpha _1\right)+\frac{1}{24} x^3 \text{$\alpha $h}^2
p_2\right)\nonumber \right.\\&-\left.
\frac{1}{2} \hbar^2 \left(K_1-\alpha _1\right) \left(x p_2^3-\frac{1}{2}
\left\{p_1 p_2^2,y\right\}-\frac{1}{2} W_2(y) p_1-y W_2'(y) p_1+\frac{3}{4}
\left\{p_2,x W_2'(y)\right\}\nonumber \right.\right.\\&-\left.\left.\frac{1}{8} \left\{p_2,y^2 W_1''(x)\right\}-\frac{1}{2}
\hbar^2 x p_2 \left(\hat{K_1}-\alpha _1\right)+\frac{1}{24} y^3 \text{$\alpha \hbar$}^2
p_1\right)\right. .
\end{flalign}  
There is just one higher order integral $X_{122}$. It is algebraically independent of $\mathcal{H}_1$ and  $\mathcal{H}_2$. Thus the system $Q_2$ with two functions $P_4$ functions is superintegrable. The coefficients $\alpha\neq0$ and can be scaled to $\alpha=-1$ to obtain the standard form of $P_4$ equation. However the value of $\alpha$ is physically significant. The potential $V(x,y)=V_1(x)+V_2(y)$ includes a harmonic oscillator term $-\frac{1}{4}\alpha (x^2+y^2)$ and hence allows a discrete spectrum (for $\alpha<0$).\\   
$Q_3:$
\begin{empheq}[box=\fbox]{align}
V_1(x)=2\hbar^2 U_1(x), \ \ \ V_2(y)=\wp(y,\hat{g_1},\hat{g_2}).
\end{empheq}
where $U_1(x)$ is given by \eqref{F-V}:
\begin{align*}
U_1{}^{(4)}(x)=&20 U_1(x) U_1''(x)+10 U_1'(x){}^2-40 U_1(x){}^3+\alpha  U_1(x)
+\gamma.
\end{align*}
This equation is a candidate for providing \textit{new} transcendents \cite{cosgrove2000higher}. The linearly independent integrals are 
\begin{flalign*}
X_{023}=&\left.\left(p_1^2+V_1(x)\right) \left( p_2^3+\frac{3}{4} \left\{p_2,V_2(y)\right\} \right)\right.=\mathcal{H}_1 X_B,\\
X_{050}=& \left. p_1^5+\frac{5}{4} \left\{p_1^3,V_1(x)\right\}+\frac{15}{16}
\left\{p_1,V_1(x){}^2\right\}+\frac{5}{16} h^2
\left\{p_1,V_1''(x)\right\}-\frac{1}{16} h^4 \alpha  p_1\right. ,&\\
X_{005}= &\left.p_2^5+\frac{5}{4} \left\{p_2^3,V_2(y)\right\}+\frac{15}{16}
\left\{p_2,V_2(y){}^2\right\}+\frac{5}{16} \hbar^2
\left\{p_2,V_2''(y)\right\}-\frac{1}{16} \hbar^2 \hat{g}_1 p_2\right. ,\\
X_{B}=& \left. p_2^3+\frac{3}{4} \left\{p_2,V_2(y)\right\} \right. .
\end{flalign*}
This case is \textit{not} superintegrable. By Theorem \ref{B-C} $X_{050}$ and $\mathcal{H}_1$ are polynomially related, as are $X_B$ and $X_{005}$ with $\mathcal{H}_2$.  Furthermore, $X_{023}$ is a product of $X_B$ and $\mathcal{H}_1$.\\
$Q_4:$
\begin{empheq}[box=\fbox]{align}
V_1(x): \text{arbitrary}, \ \ \ V_1(y)=2\hbar^2 \wp(y,\hat{g_1},\hat{g_2}).
\end{empheq}
\begin{flalign*}
X_{023}=&\left.\left(p_1^2+V_1(x)\right) \left( p_2^3+\frac{3}{4} \left\{p_2,V_2(y)\right\} \right)\right. ,\\
X_{005}=&  \left.p_2^5+\frac{5}{4} \left\{p_2^3,V_2(y)\right\}+\frac{15}{16}
\left\{p_2,V_2(y){}^2\right\}+\frac{5}{16} \hbar^2
\left\{p_2,V_2''(y)\right\}-
\frac{1}{16} \hbar^4 \hat{g}_1 p_2\right. ,&\\
X_{B}=&\left.p_2^3+\frac{3}{4} \left\{p_2,V_2(y)\right\}\right. .
\end{flalign*}
This case is ``suspect'' from the beginning since $V_1(x)$ is arbitrary. It is indeed \textit{not} superintegrable in view of Theorem \ref{B-C}. Specifically $X_B$, $X_{005}$ are algebraically dependent on $\mathcal{H}_2$ and $X_{023}$ is the product of $\mathcal{H}_1$ and $X_B$. \\
 The remaining systems $Q_5,..., Q_9$ are all superintegable and each allows just one fifth order integral. However, none of them is confining (no bound states).\\
$Q_5:$
\begin{empheq}[box=\fbox]{align}
V_1(x)=2\hbar^2 U_1(x), \ \ \ V_2(y)=2\hbar^2 P_1\left(y,\hat{B_1},\hat{B_2}\right),\ \ \ \hat{B_1}\neq0
\end{empheq}
where $U_1(x)$ satisfies (See eq.\eqref{FifIII}) 
\begin{align*}
U_1{}^{(5)}(x)=&20 U_1(x) U_1{}^{(3)}(x)+40 U_1'(x) U_1''(x)-120 U_1(x){}^2
U_1'(x)+(\alpha +x \lambda ) U_1'(x)\nonumber \\ &
+2 \lambda  U_1(x)+\omega, \ \ \ \lambda\neq0.
\end{align*}
Cosgrove showed \cite{cosgrove2000higher} that this ODE has the Painlev\'e property and may define a new transcendent, not reducible to one of the classical ones.
\begin{flalign*}
X_{023}=& \left. p_1^2 p_2^3+\frac{3}{4}
\left\{p_1^2 p_2,V_2(y)\right\}+V_1(x) p_2^3+\frac{3}{4} \left\{p_2,V_1(x)
V_2(y)\right\}+
\frac{3}{8} \left\{p_1,x V_2(y) V_2'(y)\right\}\nonumber \right.&\\&-\left.
\frac{1}{16} \hbar^2 \left\{p_1,x V_2{}^{(3)}(y)\right\}\right.\\ & +
\frac{4\hat{B_1}}{\lambda}\left(p_1^5+\frac{5}{4} \left\{p_1^3,V_1(x)\right\}+\frac{15}{16}
\left\{p_1,V_1(x){}^2\right\}+\frac{5}{16} \hbar^2
\left\{p_1,V_1''(x)\right\}\nonumber \right.\\&-\left.
\frac{1}{16} \hbar^4 \alpha  p_1\right)+ \frac{\hbar^2 \omega}{\lambda}\left(p_2^3+\frac{3}{4}
\left\{p_2,V_2(y)\right\}\right).
\end{flalign*}
$Q_6:$
\begin{empheq}[box=\fbox]{align}
V_1(x)=&\hbar^2\bigg(P_2(x,\beta)^2 + \epsilon_1 P_2'(x,\beta)\bigg), \nonumber \ \ \ \epsilon_1=\pm1\\  
V_2(y)=&\hbar^2\bigg(P_2(y,\hat{ \beta })^2 + \epsilon_2 P_2'(y,\hat{ \beta })\bigg), \ \ \ \epsilon_2=\pm1
\end{empheq}
with $\beta\neq0 ,\hat{\beta}\neq0$ and $P_2(x,\beta)=P_2(x,\beta,\mu, \delta)$ , $P_2(y,\hat{\beta})=P_2(y,\hat{\beta},\hat{\mu},\hat{\delta})$ are given by \eqref{PainlevéReduce}.
\begin{flalign*}
X_{032}=& \left.p_1^3 p_2^2+\frac{3}{4} \left\{p_1 p_2^2,V_1(x)\right\}+V_2(y)
p_1^3+\frac{3}{4} \left\{p_1,V_1(x) V_2(y)\right\}+\frac{3}{8} \left\{p_2,y V_1(x)
V_1'(x)\right\}\nonumber \right.\\&
-\left. \frac{1}{16} \hbar^2 \left\{p_2,y V_1{}^{(3)}(x)\right\}+\frac{1}{8}
\beta  \hbar^2 \left\{p_1 p_2^2,x\right\}-
\frac{1}{16} \beta 
\hbar^2 \left\{p_1,x y V_2'(y)\right\}-
\frac{3 \hbar^2 \mu}{2 \beta } \left(p_1 p_2^2+p_1 V_2(y)\nonumber \right.\right.\\&+\left.\left.
\frac{1}{4} \left\{p_2,y
V_1'(x)\right\}\right) \right.-\frac{\beta}{\hat{\beta}}\left[p_1^2
p_2^3+\frac{3}{4} \left\{p_1^2 p_2,V_2(y)\right\}+V_1(x) p_2^3+\frac{3}{4}
\left\{p_2,V_1(x) V_2(y)\right\}\nonumber \right.&\\&+\left.
\frac{3}{8} \left\{p_1,x V_2(y)
V_2'(y)\right\}-\frac{1}{16} \hbar^2 \left\{p_1,x V_2{}^{(3)}(y)\right\}+
\frac{1}{8}\hat{\beta } \hbar^2\left\{p_1^2 p_2,y\right\}
-\frac{1}{16}
\hat{\beta } \hbar^2\left\{p_2,x y V_1'(x)\right\}\nonumber \right.\\&-\left.
\frac{3 \hbar^2 \hat{\mu }}{2 \hat{\beta }}\left(p_1^2 p_2+p_2 V_1(x)+\frac{1}{4} \left\{p_1,x
V_2'(y)\right\}\right)\right].
\end{flalign*}
$Q_7:$
\begin{empheq}[box=\fbox]{align}
V_1(x)=&\hbar^2\bigg(P_2(x,\beta,\mu, \delta)^2 + \epsilon_1 P_2'(x,\beta,\mu, \delta)\bigg), \ \ \ \beta\neq0, \ \ \ \epsilon_1=\pm1 \nonumber \\
V_2(y)=&2\hbar^2 P_1(y,\hat{B_1},\hat{B_2}), \ \ \ B_1\neq0.
\end{empheq}
\begin{flalign*}
X_{023}=& \left.p_1^2 p_2^3+\frac{3}{4} \left\{p_1^2 p_2,V_2(y)\right\}+V_1(x)
p_2^3+\frac{3}{4} \left\{p_2,V_1(x) V_2(y)\right\}+\frac{3}{8} \left\{p_1,x V_2(y)
V_2'(y)\right\}\nonumber \right.&\\&-\left.
\frac{1}{16} \hbar^2 \left\{p_1,x V_2{}^{(3)}(y)\right\}-\frac{\hbar^2
	\hat{B_1}}{\beta } \left(p_1^3+\frac{3}{4} \left\{p_1,V_1(x)\right\}-\frac{3 \hbar^2 \mu
	 }{2 \beta }p_1\right)\right. .
\end{flalign*}
$Q_8:$
\begin{empheq}[box=\fbox]{align}
V_1(x)=2\hbar^2U_1(x), \ \ \ V_2(y)=2\hbar^2U_2(y),
\end{empheq}
where 
\begin{align*}
U_1{}^{(4)}(x)=&20 U_1(x) U_1''(x)+10 U_1'(x){}^2-40 U_1(x){}^3+\alpha  U_1(x)+\omega x
+\gamma,\ \ \ \omega\neq0\\
U_2{}^{(4)}(y)=&20 U_2(y) U_2''(y)+10 U_2'(y){}^2-40 U_2(y){}^3+\hat{\alpha } U_2(y)+\hat{\omega} y +\hat{\gamma }, \ \ \ \hat{\omega}\neq0.
\end{align*}

\begin{flalign*}
X_{050}=& \left.p_1^5+\frac{5}{4} \left\{p_1^3,V_1(x)\right\}+\frac{15}{16}
\left\{p_1,V_1(x){}^2\right\}+\frac{5}{16} \hbar^2
\left\{p_1,V_1''(x)\right\}-\frac{1}{16} \hbar^4 \alpha  p_1\right.&\\ 
&+\frac{\omega}{\hat{\omega}} \left(p_2^5+\frac{5}{4} \left\{p_2^3,V_2(y)\right\}+\frac{15}{16}
\left\{p_2,V_2(y){}^2\right\}+\frac{5}{16} \hbar^2
\left\{p_2,V_2''(y)\right\}-\frac{1}{16} \hbar^4 \hat{\alpha } p_2\right).
\end{flalign*}
$Q_9:$
\begin{empheq}[box=\fbox]{align}
V_1(x)=2\hbar^2 P_1(x,B_1,B_2)\ \ \ V_2(y)= 2\hbar^2 U_2(y),\ \ \ B_1\neq0
\end{empheq}
where $U_2(y)$ satisfies
\begin{align*}
U_2{}^{(4)}(y)=20 U_2(y) U_2''(y)+10 U_2'(y){}^2-40 U_2(y){}^3+\hat{\alpha } U_2(y)+\hat{\omega} y+\hat{\gamma }.
\end{align*}
\begin{flalign*}
X_{005}=&\left.p_2^5+\frac{5}{4} \left\{p_2^3,V_2(y)\right\}+\frac{15}{16}
\left\{p_2,V_2(y){}^2\right\}+\frac{5}{16} \hbar^2
\left\{p_2,V_2''(y)\right\}-\frac{1}{16} \hbar^4 \hat{\alpha } p_2\right.&\\&
 +\frac{\hat{\omega} \hbar^2}{4 B_1} \left(p_1^3+\frac{3}{4} \left\{p_1,V_1(x)\right\}\right).
\end{flalign*}
\section{Conclusion and future outlook}
We have found all doubly exotic quantum superintegrable  potentials that, in addition to the Hamiltonian, allow one second and at least one fifth-order integral of motion and are separable in Cartesian coordinates. All of these potentials are found  as solutions of equations having the Painlev\'e property. In most cases, they are expressed in terms of known transcendental functions including the six Painlev\'e transcendents. These results support the conjecture which states that all superintegrable potentials that do not satisfy any linear equation satisfy nonlinear equations having the Painlev\'e property.\\
The main result of this paper is that we have determined that among the doubly exotic systems $Q_1,...,Q_9$ of section $6$, all except $Q_3$ and $Q_4$ are superintegrable. The most interesting one is $Q_2$ that is the only one that is confining, i.e. sufficiently attractive to allow bound states. In this case we plan to use the polynomial algebra of the integrals of motion generated by $\mathcal{H}_1$, $\mathcal{H}_2$ and $X_{122}$ to calculate the energy spectrum and the wave functions \cite{daskaloyannis1991generalized,daskaloyannis2001quadratic,marquette2009superintegrability}. We also plan to construct all ``singly exotic'' potentials (e.g. when only $V_1(x)$ is exotic and $V_2(y)$ is a solution of a linear equation), specially those that are also confining. The final aim of this program is to prove the conjecture that all exotic separable superintegrable potentials have the Painlev\'e property (for arbitrary order $N$ of the additional integral).
\begin{acknowledgments}
The authors thank I. Polterovich for very helpful discussions of Burchnall-Chaundy theory, Y. Saint Aubin for many helpful comments on the manuscript and M. Sajedi for many discussions of higher order equations with the Painlev\'e property.\\
The research of P.W. was partially supported by an NSERC discovery grant. I.A. acknowledges a graduate fellowship from the Facult\'e des \'etudes sup\'erieures de l'Universit\'e de Montr\'eal.    
\end{acknowledgments}
\newpage
  \appendix
	\section{Nonlinear compatibility condition for equations  \eqref{D6}-\eqref{D8}}
	\label{appendix:graph}
		\resizebox{.95\linewidth}{!}{
		\begin{minipage}{\linewidth}
	\begin{align}
&\frac{1}{2} f_{12}  V_{xxx}+\left(\frac{3}{2} f_{02}-f_{22} \right)
V_{xxy}+\left(-f_{12} +\frac{3}{2} f_{32} \right) V_{xyy}+\frac{1}{2} f_{22}V_{yyy} 
+\left(-f_{22}{}^{(0,1)} +f_{12}{}^{(1,0)} \right) V_{xx}\nonumber \\ &+\left(-f_{12}{}^{(0,1)} +3
f_{32}{}^{(0,1)} +3 f_{02}{}^{(1,0)} -f_{22}{}^{(1,0)} \right) V_{xy}+ \left(f_{22}{}^{(0,1)} -f_{12}{}^{(1,0)} \right)V_{yy} \nonumber \\ &
+\left(\frac{3}{2}
f_{32}{}^{(0,2)} -f_{22}{}^{(1,1)} +\frac{1}{2} f_{12}{}^{(2,0)} \right) V_x  + \left(\frac{1}{2} f_{22}{}^{(0,2)} -f_{12}{}^{(1,1)} +\frac{3}{2} f_{02}{}^{(2,0)} \right)V_{y} \nonumber \\ &
+h^2 \left[\left(-\frac{1}{4} f_{10} -\frac{1}{8} f_{30} \right) V_{xxxxx} +\left(\frac{-5}{4} f_{00} +\frac{1}{8}f_{20} +\frac{1}{2} f_{40} \right) V_{xxxxy}+\left(\frac{1}{4} f_{10} -\frac{5}{4} f_{50} \right) V_{xxxyy}\nonumber \right.\\&+\left.
\left(\frac{-5}{4} f_{00} + \frac{1}{4} f_{40} \right) V_{xxyyy}+
\left(\frac{1}{2} f_{10} +\frac{1}{8}f_{30} -\frac{5}{4} f_{50} \right) V_{xyyyy}+\left(-\frac{1}{8} f_{20} -\frac{1}{4} f_{40} \right)
V_{yyyyy} \nonumber \right.\\&+\left.
\left(\frac{1}{2} f_{20}{}^{(0,1)} +\frac{1}{2}
f_{40}{}^{(0,1)} -f_{10}{}^{(1,0)}
-\frac{1}{4} f_{30}{}^{(1,0)} \right) V_{xxxx}\nonumber \right.\\&+\left.
\left(f_{10}{}^{(0,1)} -\frac{3}{4} f_{30}{}^{(0,1)} -\frac{5}{2}f_{50}{}^{(0,1)} -5 f_{00}{}^{(1,0)} +\frac{3}{4} f_{20}{}^{(1,0)} +
\frac{1}{2} f_{40}{}^{(1,0)} \right) V_{xxxy} \nonumber \right.\\&+\left.
\left(\frac{-3}{4}f_{20}{}^{(0,1)} +\frac{3}{2}f_{40}{}^{(0,1)} +\frac{3}{2} f_{10}{}^{(1,0)} -\frac{3}{4} f_{30}{}^{(1,0)} \right) V_{xxyy}\nonumber \right.\\&+\left.
 \left(\frac{1}{2} f_{10}{}^{(0,1)} +\frac{3}{4} f_{30}{}^{(0,1)} -5 f_{50}{}^{(0,1)} -\frac{5}{2}
f_{00}{}^{(1,0)} -\frac{3}{4} f_{20}{}^{(1,0)} +f_{40}{}^{(1,0)} \right) V_{xyyy}
\nonumber \right.\\&+\left.
\left(-\frac{1}{4} f_{20}{}^{(0,1)} -f_{40}{}^{(0,1)} +\frac{1}{2} f_{10}{}^{(1,0)} +\frac{1}{2}
f_{30}{}^{(1,0)} \right)V_{yyyy} \nonumber \right.\\&+\left.
\left(\frac{-3}{4}f_{10}{}^{(0,2)} -\frac{3}{4}
f_{30}{}^{(0,2)} -\frac{5}{4} f_{50}{}^{(0,2)} +\frac{3}{4} f_{20}{}^{(1,1)} +\frac{1}{2} f_{40}{}^{(1,1)} -\frac{3}{2}f_{10}{}^{(2,0)} -\frac{1}{2} f_{30}{}^{(2,0)} \right) V_{xxx} \nonumber \right.\\&+\left.
\left(\frac{-15}{4} f_{00}{}^{(0,2)} +\frac{9}{4} f_{40}{}^{(0,2)} +\frac{3}{2} f_{10}{}^{(1,1)} -\frac{3}{4} f_{30}{}^{(1,1)} -\frac{15}{2} f_{00}{}^{(2,0)} +\frac{3}{4}f_{20}{}^{(2,0)}+
\frac{3}{2} f_{40}{}^{(2,0)} \right) V_{xxy}\nonumber \right.\\&+\left.
\left(\frac{3}{2} f_{10}{}^{(0,2)} +\frac{3}{4}f_{30}{}^{(0,2)} -\frac{15}{2} f_{50}{}^{(0,2)} -\frac{3}{4} f_{20}{}^{(1,1)} +\frac{3}{2} f_{40}{}^{(1,1)} +\frac{9}{4}f_{10}{}^{(2,0)}-
\frac{15}{4} f_{50}{}^{(2,0)} \right)V_{xyy} \nonumber \right.\\&+\left.
\left(-\frac{1}{2} f_{20}{}^{(0,2)} -\frac{3}{2}f_{40}{}^{(0,2)} +\frac{1}{2} f_{10}{}^{(1,1)} +\frac{3}{4} f_{30}{}^{(1,1)} -\frac{5}{4} f_{00}{}^{(2,0)} -\frac{3}{4}f_{20}{}^{(2,0)}
-\frac{3}{4} f_{40}{}^{(2,0)} \right) V_{yyy}  \nonumber \right.\\&+\left.
\left(\frac{3}{4}
f_{20}{}^{(0,3)} +f_{40}{}^{(0,3)} -\frac{3}{2} f_{10}{}^{(1,2)} +\frac{3}{2}f_{40}{}^{(2,1)} -f_{10}{}^{(3,0)} -\frac{3}{4} f_{30}{}^{(3,0)} \right)V_{xx}  
\nonumber \right.\\&+\left.
 \left(\frac{3}{2}f_{10}{}^{(0,3)} -\frac{1}{4} f_{30}{}^{(0,3)}-5 f_{50}{}^{(0,3)} -\frac{15}{2} f_{00}{}^{(1,2)} +\frac{3}{4}f_{20}{}^{(1,2)} +\frac{3}{4} f_{30}{}^{(2,1)} -\frac{15}{2} f_{50}{}^{(2,1)}-
5 f_{00}{}^{(3,0)} \nonumber\right. \right.\\&+\left. \left.-\frac{1}{4}f_{20}{}^{(3,0)} +\frac{3}{2} f_{40}{}^{(3,0)} \right)V_{xy} 
+ \left(\frac{-3}{4}f_{20}{}^{(0,3)} -f_{40}{}^{(0,3)} +\frac{3}{2} f_{10}{}^{(1,2)} -\frac{3}{2}
f_{40}{}^{(2,1)} +f_{10}{}^{(3,0)} \nonumber \right. \right.\\&+\left. \left.\frac{3}{4} f_{30}{}^{(3,0)} \right)V_{yy}  
+ \left(\frac{-3}{8}f_{30}{}^{(0,4)}-
\frac{5}{4} f_{50}{}^{(0,4)} +\frac{3}{4} f_{20}{}^{(1,3)} 
-\frac{1}{2} f_{40}{}^{(1,3)} -\frac{3}{4}f_{10}{}^{(2,2)} +\frac{3}{4} f_{30}{}^{(2,2)} \nonumber \right. \right.\\&-\left. \left. \frac{15}{4} f_{50}{}^{(2,2)} -\frac{1}{4} f_{20}{}^{(3,1)} +\frac{3}{2}f_{40}{}^{(3,1)}  \right)V_{x} + \left(\frac{-3}{8}f_{20}{}^{(0,4)} -\frac{1}{4} f_{40}{}^{(0,4)} +\frac{3}{2} f_{10}{}^{(1,3)} -\frac{1}{4} f_{30}{}^{(1,3)} \nonumber \right. \right.\\&-\left. \left. 
\frac{15}{4}f_{00}{}^{(2,2)}+
\frac{3}{4} f_{20}{}^{(2,2)} -\frac{3}{4} f_{40}{}^{(2,2)} -\frac{1}{2} f_{10}{}^{(3,1)} +\frac{3}{4}f_{30}{}^{(3,1)} -\frac{5}{4} f_{00}{}^{(4,0)} -\frac{3}{8} f_{20}{}^{(4,0)} \right)V_{y} \nonumber \right.\\&-\left. \frac{1}{4} f_{10}{}^{(4,0)} -\frac{3}{8} f_{30}{}^{(4,0)} \right]=0. \label{appen}				
	\end{align}
		\end{minipage}
}
\newpage
\bibliography{Bibliography_article}
\end{document}